\documentclass[journal,10pt]{IEEEtran}

\usepackage{amsmath,amsfonts,amssymb,amstext}

\usepackage{color}

\usepackage{multicol}

\usepackage{array}

\usepackage[hyphens]{url}
\usepackage{hyperref}

\usepackage[T1]{fontenc}
\usepackage{lmodern}

\usepackage[normalem]{ulem} % provides \sout

\definecolor{gris}{gray}{0.5}

\newcommand{\forget}[1]{}

\newcommand{\ran}{\xleftarrow{~~}}
\newcommand{\bt}[1]{{\left( #1 \right)}}
\newcommand{\sbt}[1]{{\left[ #1 \right]}}
\newcommand{\abt}[1]{{\left| #1 \right|}}
\newcommand{\setd}[1]{{\left\{ #1 \right\}}}

\newcommand{\A}{{\cal A}}
\newcommand{\B}{{\cal B}}

\newcommand{\CH}{{\cal H}}

\newcommand{\rand}{\leftarrow}

\newcommand{\pair}{{\sf \hat e}}
\newcommand{\adv}{{\sf Adv}}
\newcommand{\prob}{{\sf Pr}}

\usepackage{subfigure}
\usepackage{multicol}
\usepackage{graphicx}
\graphicspath{ {slides/images/} }

\usepackage{paralist}

\usepackage{enumitem}

\usepackage{cleveref}

\usepackage{times}
\usepackage{textcomp}

\usepackage{caption}
\usepackage{array,multirow}

\usepackage{graphics}

\usepackage{mathtools}

\usepackage[font=scriptsize,labelfont=bf]{caption}

\usepackage{relsize}

\newcommand{\eid}{{\sf et}}

\newcommand{\sn}{{\sf sn}}

\newcommand{\ukg}{{\sf UKg}}

\newcommand{\epk}{{\it epk}}
\newcommand{\esig}{{\sf ESign}}
\newcommand{\ever}{{\sf EVer}}

\newcommand{\hs}{{\CH_1}}
\newcommand{\hsh}{{\CH_2}}

\newcommand{\gpk}{{\sf gpk}}
\newcommand{\mik}{{\sf mik}}
\newcommand{\mok}{{\sf mok}}
\newcommand{\usk}{{\sf usk}}
\newcommand{\upk}{{\sf upk}}
\newcommand{\gsk}{{\sf gsk}}

\newcommand{\setup}{{\sf GSet}}

\newcommand{\join}{{\sf Join}}
\newcommand{\issue}{{\sf Issue}}
\newcommand{\gsig}{{\sf GSign}}
\newcommand{\gver}{{\sf GVer}}
\newcommand{\open}{{\sf Open}}
\newcommand{\judge}{{\sf Judge}}
\newcommand{\link}{{\sf Link}}

\newcommand{\reg}{{\sf reg}}

\newcommand{\negl}{{\sf negl}}

\newcommand{\name}{{AEE}}

\newcommand{\gcor}{{\sf Exp}_{\A}^{\sf corr}}
\newcommand{\gany}{{\sf Exp}_{\A}^{{\sf anony}}}
\newcommand{\gnfg}{{\sf Exp}_{\A}^{\sf gsig\mbox{-}non\mbox{-}frame}}
\newcommand{\gnfe}{{\sf Exp}_{\A}^{\sf esig\mbox{-}unforg}}
\newcommand{\gtr}{{\sf Exp}_{\A}^{\sf trace}}
\newcommand{\gel}{{\sf Exp}_{\A}^{\sf link\mbox{-}unforg}}
\newcommand{\glk}{{\sf Exp}_{\A}^{\sf e\mbox{-}link}}

\newcommand{\advc}{{\sf Adv}^{\sf corr}_{\A}}
\newcommand{\adva}{{\sf Adv}^{\sf anony}_{\A}}
\newcommand{\advnfg}{{\sf Adv}^{\sf gsig\mbox{-}non\mbox{-}frame}_{\A}}
\newcommand{\advnfe}{{\sf Adv}^{\sf esig\mbox{-}unforg}_{\A}}
\newcommand{\advtr}{{\sf Adv}^{\sf trace}_{\A}}
\newcommand{\advlk}{{\sf Adv}^{\sf e\mbox{-}link}_{\A}}
\newcommand{\advel}{{\sf Adv}^{\sf link\mbox{-}unforg}_{\A}}

\newcommand{\elgs}{{\cal AEE}}

\newcommand{\hu}{{\sf HU}}
\newcommand{\cu}{{\sf CU}}
\newcommand{\bu}{{\sf BU}}

\newcommand{\ch}{{\sf Ch}_b}
\newcommand{\isl}{{\sf SigL}}
\newcommand{\cl}{{\sf ChL}}

\newcommand{\pr}{{\sf Pr}}

\newcommand{\mul}{{\it mul}}
\newcommand{\iexp}{{\it exp}}

\newcommand{\addu}{{\sf AddU}}

\newcommand{\rreg}{{\sf RReg}}
\newcommand{\wreg}{{\sf WReg}}
\newcommand{\sndu}{{\sf SndToU}}

\newcommand{\ogsk}{{\sf USK}}

\newcommand{\fail}{{\sf fail}}

\newtheorem{definition}{Definition}
\newtheorem{theorem}{Theorem}
\newenvironment{proof}[1]{{\medskip\noindent\bf Proof of #1:}}{}%\hfill$\square$}

\begin{document}

\title{A Novel Computationally Efficient Group Signature for Anonymous and Secure V2X Communications}

\author{Jia Liu, Liqun Chen,~\IEEEmembership{Member,~IEEE}, Mehrdad Dianati,~\IEEEmembership{Senior Member,~IEEE}, Carsten Maple and Yan Yan

\thanks{Jia Liu is with the Institute for Communication Systems (ICS), 5G Innovation Centre (5GIC), University of Surrey, United Kingdom, GU2 7XH, UK (e-mail: j.liu@surrey.ac.uk)}
\thanks{Liqun Chen is with Department of Computer Science,  University of Surrey, Guildford, GU2 7XH, UK (e-mail: liqun.chen@surrey.ac.uk) }
\thanks{Mehrdad Dianati is with School of Electronics, Electrical and Computer Science, Queen's University of Belfast, Northern Ireland, BT7 1NN (e-mail: M.Dianati@qub.ac.uk)}
\thanks{Carsten Maple is with the Cyber Security Centre, University of Warwick, Warwick CV4 7AL, U.K. (e-mail: carsten.maple@warwick.ac.uk )}
\thanks{Yan Yan is with Department of Computer Science, University of Bristol, BS8 1UB, UK (e-mail: y.yan@bristol.ac.uk)}
}

%   \frontmatter

%\pagestyle{\markright{VeriCloud} }

% set page number position

\maketitle

\begin{abstract}
The use of vehicle-to-everything (V2X) communication is expected to significantly improve road safety and traffic management. We propose a novel efficient protocol, called {\name} protocol, for protecting data authenticity and user privacy in V2X applications.
%Data authenticity and user privacy in V2X applications can usually be achieved using anonymous signatures. However, existing anonymous signatures designed for vehicular communication are either not efficient enough for the delay sensitive V2X applications, or lack of a certain  level of linkability for retaining accountability.
Our protocol provides event-based likability, which enables messages from a subject vehicle to be linked to a specific event in order to prevent Sybil attacks. Messages on different events are unlinkable to preserve the long-term privacy of vehicles. Moreover, our protocol introduces a new method for generating temporary public keys to reduce computing and communication overheads. Such a temporary public key is bound with a certain event and is automatically revoked when the event is over. % Our scheme firstly generates a group signature with an event-linking token which is a unique token on each event for each user. Then,  to reduce computing and transmission overheads, the event-linking token is used as a self-certified temporary public key for generating traditional signatures to authenticate the subsequent messages. 
We describe how to apply our protocol in vehicular communications using two exemplar use cases. %: intersection management and cooperative awareness messages. 
To further reduce the real-time computational complexity, our protocol enables us to decompose the cryptographic operations into offline processes for complex operations and real-time processes for fast computations. %we propose an offline computation solution for generating group signatures that requires complex computations to further reduce the processing time of the proposed scheme.

%Our concept of event linkability generalises the previous notion of short-term linkability by enabling flexibilities on how the vehicles are linked in different situations. 
\end{abstract}

\section{Introduction}
%\paragraph{Vehicular networks} Approximately 25,900 people were killed in road accidents in the EU countries in 2014, according to the latest traffic safety statistics (Traffic Safety Basic Facts 2016, 2016); National Highway Traffic Safety Administration (NHTSA) reports the number of people killed on the road in the U.S. soared 7.2\% to 35,092 in 2015. Road Safety remains a major societal issue. Vehicular Ad hoc Networks (VANETs) has tremendous potential to improve road safety and traffic management and provide Internet access on highway.

Future vehicles are envisaged to use V2X (vehicle to everything) communications for a variety of safety, driving efficiency and infotainment applications. For example,  vehicles use V2X communication technology to periodically (e.g., every 0.1s \cite{etsi2014}) communicate their status information, such as position, speed, heading and acceleration, to surrounding vehicles and infrastructures. 
%A vehicle's status information is updated periodically, for example every 0.1s as specified in the European standard \cite{etsi2014}. Through the communications, 
Vehicles can also be informed of crucial traffic information such as accidents, ice, fog, and rain.
Vehicle awareness of its environment is increasingly considered to improve collision avoidance and reduce fatalities and injury severity. 
In this work, we shall consider how to secure two representative  safety applications that rely on V2X communications: intersection management and cooperative awareness messages. Intersection management uses the communication between an intersection controller and the nearby vehicles to coordinate vehicles to cross the intersection safely. Cooperative awareness messages are broadcast periodically from each vehicle to inform other vehicles about their presence and status.
%While on the road, vehicles by,for instance, employing specialized embedded sensors may notice the presence of certaincritical trafﬁc conditions, such as trafﬁc jams, accidents, etc., or road conditions, such as ice, fog, rain etc., that increase the risk of accidents.
 % can be informed of crucial traffic information such as treacherous road conditions and accident sites through the communications. 

%VANETs applications include crash warning, collision avoidance, cooperative driving, traffic optimisation, dissemination of weather information, etc.

Securing V2X communication is an indispensable prerequisite for acceptance of the applications enabled by such technologies. %Messages broadcasted in the vehicular networks should be authentic.
Harmful information from a malicious node can jeopardize the safety of target vehicles and endanger others in the vicinity. Therefore, on the one hand, it is a basic requirement to guarantee that information comes from a trusted source and has not been tampered with during transmission. Also, there is a need for tracking malicious nodes and behaviours. On the other hand, protecting vehicle privacy is also of great importance since communications in vehicular networks can be easily abused for vehicle tracking and compromising the privacy of the users. %The consequences of a security breach in vehicular communications can be critical and fatal. The most essential security attributes in vehicular communications  are to ensure that the information comes from a trusted source (e.g., authentication, non-repudiation) and has not been tampered with during transmission (e.g., integrity). Securing vehicular communications not only determines the responsibility of vehicles but also maintain the vehicle user's privacy. 
For example, %as a vehicle's trajectory information accumulates over time, 
the locations visited by a vehicle enable inference and profiling of the personal interests of its user. %Such attacks are not difficult since a private vehicle is typically only driven by few individuals and the adversary may track the users until their real identities are discovered. 
To this end, achieving security and privacy requirements at the same time is a challenging objective. % contradict with each other. % in many technical aspects. 

A feasible balance between security and privacy in vehicular networks is the short-term linkability originally proposed in \cite{DBLP:conf/secon/StuderSBP09}. Short-term linkability allows tracking of the movement of a vehicle in a short period of time in order to thwart Sybil attacks (i.e., one vehicle claims to be multiple vehicles) as early as possible while preserving long-term privacy. This is a crucial property for many applications in vehicular networks, such as live traffic map generation, intersection management, and cooperative position.  For example, a malicious vehicle may impersonate multiple vehicles and fake traffic congestion to deceive other vehicles to divert the traffic.

Beyond the security features, in practice, computational efficiency is a big obstacle for deploying any security mechanism into vehicular networks. The exchange of vital information for safety-related V2X applications is very delay sensitive. In high-speed cooperative driving scenarios,  the communication overhead of each packet and the computation latency at each vehicle must be very low to ensure the information exchange is effective. For example, the data processing time for a safety message should be less than 50ms according to the European standard \cite{etsi2014}.  To simultaneously achieve these seemingly contradicting requirements without compromising the functionality of V2X applications remain a challenge.

%which make them unsuitable for delay sensitive applications such as exchange of vital information for safety related V2X applications.

%Simultaneously achieving security, privacy and efficiency remains a challenge in securing vehicular communications due to their seemly contradictory requirements. 

{\medskip\noindent\bf Our contributions.}
This paper proposes the {\name} protocol, an efficient, secure and privacy-enhancing protocol for V2X applications.
%This paper introduces the concept of event linkability for securing V2X applications  and  an anonymous  signature scheme with event-based linkability and revocation, referred to as $\elgs$. Tailored for V2X applications, the proposed $\elgs$ scheme 
%Our protocol aims to achieve all the security and privacy properties along with high efficiency.  In this scheme, 
Our  protocol provides event-based linkability to prevent Sybil attacks. An event is uniquely identified using, for example, a timestamp, a location name, a random number, or an incident summary, depending on the use case. The proposed protocol first generates a modified group signature to include an event-linking token which is a unique token on each event for each user. Therefore, multiple signatures  produced by the same user can be linked together on the basis of the event. However, messages signed by the same entity on different events are unlikable, which ensures the user's long-term privacy. This modified group signature scheme has its independent interests. To reduce computing and transmission overheads introduced in group signatures, we re-use the event-linking token as a self-certified temporary public key for generating simple and efficient traditional signatures called {\em event signatures} to authenticate the subsequent messages. The group signature acts as a certificate for the temporary public key. Since each user can only create one temporary public key for each event, when the event is over, the temporary public key is automatically revoked. To ensure the desired level of privacy,  the event has to be updated  in an appropriate time frame.

%Our concept of event linkability generalises the previous notion of short-term linkability \cite{Golle:2004:DCM:1023875.1023881} by enabling flexibilities on how the vehicles are linked, instead of the  vague length of time. 
To demonstrate the applicability of the proposed scheme,   we illustrate how to apply our scheme to V2X communications by considering two use cases: intersection management \cite{7313162,6728435} and cooperative awareness messages \cite{etsi2014}.  When future events are predictable, we devise an offline computing mechanism to generate the time-consuming group signatures in advance, which significantly reduces the computational delay of the cryptographic operations.
Security properties of our scheme are formalised as anonymity, traceability, event linkability and unforgeability in a model for dynamic group signatures.
We prove these properties using the random oracle model. 
%\begin{itemize}[leftmargin=*]
%\item {\em Anonymity} ensures that the outgoing messages do not reveal the identity of the vehicle who produced them. Signatures on different events from one source cannot be linked together. 
%\item {\em Traceability} enables a designated trusted authority to revoke the anonymity of the message signer  when there is a dispute. This aims to identify and punish the malicious notes in case of liability investigation.
%\item {\em Event linkability} links messages involved in the same event from the same source. Event linkability prevents Sybil attacks in an early stage, compared to the traceability. A vehicle with a valid credential is not able to produce multiple messages on the same event that appear to be from different vehicles.
%\item {\em Unforgeability} ensures that a valid signature, including group signature and event signature, cannot be attributed to an honest member unless this member does produce it. It also guarantees that the link among multiple messages cannot be forged. %, that is, messages from different sources will not be linked while messages on the same event from the same source will be correctly linked.
%\end{itemize}

%We prove these properties using the random oracle model. 
%Compared to the standard model \cite{DBLP:conf/asiacrypt/Groth07}, the random oracle model leads to much more light-weight implementation of group signature schemes.

{\medskip\noindent\bf Outline.} The rest of this paper is organised as follows:  
\Cref{sec:relatedwork} describes and summarises the existing related work.
\Cref{sec:crypto} describes the cryptographic preliminaries and  computational assumptions.
\Cref{sec:model} presents the model and security goals for our protocol.
\Cref{sec:construction} presents the construction of our protocol.
\Cref{sec:app} illustrates how to use our scheme to secure applications in V2X communications. 
\Cref{sec:implementation} evaluates performance of our protocol. 
\Cref{sec:secanalysis} analyses security properties for our protocol.  
The paper concludes in \Cref{sec:conclusion}.

\section{Related work.}\label{sec:relatedwork}
There is a large body of literature on anonymous authentication schemes.
Here we will focus on the research that is tailored to the vehicular networks.
Short group signature scheme was originally proposed in  
\cite{Boneh2004}  to provide anonymous authentications for each message broadcast in vehicular networks. Group signatures achieve authenticity, data integrity, anonymity, and accountability, while getting rid of the heavy overhead for handling numerous public key certificates in the PKI-based pseudonym schemes, e.g., \cite{Gollan02digitalsignatures,5698239}. A group signature scheme enables members of a group to sign messages on behalf of the group. Signatures are verified using a single group public key, and thus they do not reveal the identity of the signer. %Furthermore, it is not possible to decide whether two signatures have been issued by the same group member, which effectively prevents a user from being tracked.

However, it is not effective to directly apply a group signature scheme to applications in vehicular networks. This is because most of the group signature schemes, e.g., \cite{Boneh2004,Brickell:2004:DAA:1030083.1030103,Chen2010,5290062,5719272,cryptoeprint:2014:926,7006753,7047924}, are based on cryptographic paring operations which are known to be computationally expensive. This makes group signatures unsuitable for direct and frequent authentication in delay-sensitive safety applications in V2X communications. %Although currently there is no agreement about a vehicle's on-board hardware capabilities, the vehicle's computing power is restricted by the fact that it is powered by the fuel and battery.  %the size of a short group signature scheme can vary from 6 elements to 50 elements, e.g.,, depending on the supported functions and security analysis (e.g., random oracle model or standard model).
Several papers \cite{Calandriello:2007:ERP:1287748.1287752,4199150,Lu_ecpp:efficient,DBLP:conf/secon/StuderSBP09} propose hybrid solutions that generate a new temporary public/secret key pair and authenticate them with a group signature and then sign messages with a traditional digital signature scheme using the temporary private key.
However, the aforementioned  approaches suffer from several shortcomings. First, negotiating, creating and transmitting new temporary keys introduces extra overhead.  It is also a  non-trivial work if a verifier requires proof of the ownership of the temporary private key. Second,  short-term linkability   is not securely achieved.  In the existing work \cite{Calandriello:2007:ERP:1287748.1287752,4199150,Lu_ecpp:efficient,DBLP:conf/secon/StuderSBP09}, short-term linkability is achieved   by fixing the randoms used in group signatures or the temporary public key where a malicious user does not have to follow through and can break the linkability.

Threshold authentication  is proposed in \cite{DVDFJSFVA,5719272,7047924,5290062} for VANET communications  where a message is viewed as trustworthy only after it has been endorsed by a certain number of vehicles. Since all the vehicles have to sign exactly the same message to increase the trustworthiness of the message, this method has several limitations. First of all, a safety message needs to be associated with a timestamp to ensure its effectiveness, but the timestamp may vary from vehicle to vehicle, and the messages from different vehicles cannot be exactly the same. Secondly, the threshold method does not apply to the use cases considered in our paper where a vehicle measures and signs its own kinematic information, which cannot be endorsed by any other vehicles.

A time-dependent linking system \cite{cryptoeprint:2014:926} is proposed for vehicle-to-infrastructure communications where a token-generation unit broadcasts a time-token periodically. In comparison, the event in our scheme can be flexibly chosen and customised according to the regions and the real-time traffic conditions. Our scheme enables us to design an offline computation mechanism for the generation of group signatures when the future event is predictable or deterministic. Moreover, the linking token in \cite{cryptoeprint:2014:926} cannot be reused as a temporary public key, as our scheme can do.

Moreover, the schemes in \cite{5719272,cryptoeprint:2014:926,5290062} do not support efficient opening, and the computational time for tracing a malicious vehicle is linear in the total number of vehicles. We stress that an efficient opening mechanism is indispensable for ensuring the accountability of vehicles and punishing malicious behaviours. 
The notion of linkability is not formalised in  \cite{7047924,5290062}. A notion of linking soundness is defined in \cite{cryptoeprint:2014:926}, but linking soundness only ensures that the attacker cannot forge a link among messages that are not expected to be linked, and this property can be satisfied by an arbitrary group signature scheme. A crucial aspect of linkability for preventing Sybil attacks is that the users cannot bypass the linkability of the messages that are supposed to be linked. We propose a framework which formalises anonymity, traceability, event linkability and unforgeability. The scheme in \cite{7047924} is very time-consuming, and the security proofs are based on unusual assumptions.

\section{Preliminaries}\label{sec:crypto}
Our protocol makes use of cryptographic bilinear maps. In this section, we review the definitions of cryptographic bilinear maps and describe the computational assumptions for the security of our scheme.

{\em Bilinear Maps.}
Let $\mathbb{G}_1, \mathbb{G}_2$ and $\mathbb{G}_T$ be multiplicative groups of prime order $p$. A function $\pair: \mathbb{G}_1\times \mathbb{G}_2 \rightarrow \mathbb{G}_T$ is a bilinear map if it satisfies the following three properties:
\begin{enumerate}[leftmargin=0.2in]
  \item Bilinear: $\pair(g^a, h^b) = \pair(g,h)^{ab}$ for all $g\in \mathbb{G}_1, h\in \mathbb{G}_2$ and $a,b\in \mathbb{Z}^*_p$.
  \item Non-degenerate: there exists $g\in \mathbb{G}_1, h\in \mathbb{G}_2$ such that $\pair(g,h)\neq 1$.
      \item Computable: there exists an efficient algorithm to compute $\pair(g,h)$ for all $g\in \mathbb{G}_1, h\in \mathbb{G}_2$.
\end{enumerate}

%\subsection{Computational assumptions}
\begin{definition}[Discrete Logarithm (DL) assumption]
The DL assumption holds if, for all PPT adversaries $\A$,
  \[
  \adv_{{\cal A}}^{\sf DL}(1^\lambda) = \prob\sbt{\A(\mathbb{G}, q, g, g^x) = x: x \ran \mathbb{Z}^*_p} 
  \] is negligible in $\lambda$. % where in each case the probabilities are taken over the experiment in which $(\mathbb{G},q,g)\ran\G(1^\lambda)$.
\end{definition}

\begin{definition}[Decision Diffie-Hellman (DDH) assumption]
  DDH assumption holds if for all PPT adversaries $\cal A$,
  \[
  \begin{split}
  \adv_{{\cal A}}^{\sf DDH}&(1^\lambda) = \\
 & \abt{
  \begin{split}
 & \prob\sbt{\A(\mathbb{G}, q, g, g^a, g^b, g^{ab}) = 1} -\\ &~~~~ \prob\sbt{\A(\mathbb{G}, q, g, g^a, g^b, g^{c}) = 1}: a,b,c \ran \mathbb{Z}^*_p
  \end{split}
  }
    \end{split}
  \] is negligible in $\lambda$. % where in each case the probabilities are taken over the experiment in which $(\mathbb{G},q,g)\ran\G(1^\lambda)$ and $a,b,c\in\mathbb{Z}^*_p$ are randomly chosen.
\end{definition}

\begin{definition}[EXternal Diffie-Hellman (XDH) Assumption]
Given groups $\mathbb{G}_1, \mathbb{G}_2, \mathbb{G}_T$ associated with a bilinear pairing $\pair: \mathbb{G}_1\times \mathbb{G}_2 \rightarrow \mathbb{G}_T$. The XDH assumption holds if the DDH problem is hard in $\mathbb{G}_1$ but easy in $\mathbb{G}_2$.
\end{definition}

\begin{definition}[$q$-SDH Assumption]
The $q$-Strong Diffie-Hellman (SDH) assumption is: given two multiplicative groups $\mathbb{G}_1$ and $\mathbb{G}_2$ of prime order $p$ with generators $g_1$ for $\mathbb{G}_1$ and $g_2$ for $\mathbb{G}_2$. For any PPT adversary $\A$, the advantage
\[
\adv_{{\cal A}}^{\sf q\mbox{-}SDH}(1^\lambda) = \pr[\A(g_1,g_1^\gamma,\cdots, g_1^{\gamma^q}, g_2, g_2^\gamma) = (g_1^{\frac{1}{\gamma+x}},x)]
\] is negligible in $\lambda$.
\end{definition}

\section{Models and security goals}\label{sec:model}
In this section, we describe the system model, clarify the security assumptions and specify the security and privacy requirements.

\subsection{System model}
Entities involved in our model are: 
\begin{itemize}
\item On-board Unit (OBU): this represents the hardware and software component that is part of a vehicle, such as GPS receiver and kinematic sensors. An OBU has wireless interfaces such as DSRC, Bluetooth or LTE, through which the OBU can broadcast its status information and surrounding traffic information to its nearby OBUs.
\item Issuer: this is an authority that authorises which OBU can join the group to become a legitimate member. The issuer creates membership credentials for legitimate OBUs, which can be used to anonymously sign all the outgoing messages from these OBUs. In the context of the vehicular network, 
%as illustrated in \Cref{fig:overview}, 
the issuer could be a transport authority, such as Driver \& Vehicle Licensing Agency (DVLA) in the United Kingdom. 
\item Opener: this is an authority  that can revoke the anonymity of a misbehaved OBU  by opening its signatures. An opener could be a law enforcement agency or a judge. %, or big organisations.
\item Roadside Unit (RSU):  RSU is installed with hardware and software components, such as an intersection scheduling algorithm.
RSU is distributed along the roadside, typically at the intersection area. RSU manages the OBUs that come within their wireless communication range and connects to the backbone networks. 
\end{itemize}

\subsection{Security assumptions}
We assume that OBUs are untrustworthy but tamper-evident. Tamper-evident describes the ability to detect and keep indisputable evidence when an adversary attempts to modify or break  the hardware and software components. This is a common assumption (e.g., \cite{5719272,7580758}) and can be achieved using hardware and software security mechanisms such as the Trusted Platform Module (TPM) for automotive \cite{TPMAuto}. 
Each OBU is assumed to be equipped with a tamper-evident black box which provides secure storage for cryptographic keys and performs cryptographic operations, e.g., generating public/secret keys and creating and verifying a signature. The black box is also pre-loaded with public information, such as system parameters and authorities' public keys. This information can later be updated in a trustworthy manner using signed updates. %A trusted authority can be a transport authority or a vehicle manufacturer. 
An OBU registers with the issuer and obtains an anonymous credential which will be securely stored in the black box. Further discussions on the V2X applications can be found in \Cref{sec:app}.

% Each black box is assumed to be associated with a pair of public/private key certified by a transport authority or car manufacturer. , which enable the black box to prove its identity and legitimacy.

%A vehicle is a user whose public/secret key can be certified by a transport authority or car manufacturer and is stored in the black box prior to any system setup. %The certification of a vehicle's public key relies on the typical public key infrastructure (PKI) and the details are omitted here. 

%V2X communications mainly consist of vehicle-to-vehicle (V2V) communication and vehicle-to-infrastructure (V2I) communication. 

The assumptions on the capabilities  of an adversary can vary for different security properties. Certain security properties, such as unforgeability, can be preserved even when an adversary learns  the secret keys of authorities. The formal definitions of an adversary's capabilities for each security goal are presented in \Cref{fig:games} in the appendix. Roughly speaking, we consider the common threat model where an adversary can eavesdrop the wireless communication to collect, inject and modify messages. 
%An adversary may be an external intruder, and may also be an authenticated internal member who has a valid credential. 
An adversary may possess some legitimate OBUs, break the devices to retrieve the stored cryptographic keys or modify the cryptographic algorithms.
  However, we assume the adversary cannot break the underlying cryptographic assumptions given in the following \Cref{sec:crypto}.

\subsection{Correctness and security goals}
The correctness and security properties of our {\name} protocol are described below. Their formal definitions are defined using experiments involving an adversary and a challenger  and can be found in  Figure \ref{fig:games} in the appendix.

\paragraph{Correctness} The correctness guarantees that 
\begin{inparaenum}
\item signatures produced by honest OBUs are accepted by the verification algorithms; 
\item the honest opener can identify the signer of such signatures;
\item multiples signatures signed by the same honest OBU for the same event can be correctly linked. 
\end{inparaenum}

\paragraph{Anonymity} This requires that signatures do not reveal any information about the identity of the OBUs who produced them. 
Signatures produced by one signer but on different events should be unlinkable.

\paragraph{Traceability}  This ensures that the adversary cannot produce a signature that cannot be traced to a legitimate OBU. 
A designated trusted authority can revoke the anonymity of the message signer when there is a dispute. It aims to identify and punish malicious OBUs in case of liability investigation. 
\paragraph{Event linkability}  This links signature involved in the same event signed by the same OBU, while signatures signed by the different OBUs or on different events cannot be linked.  This implies that an OBU cannot produce unlinkable signatures for the same event. Event linkability prevents Sybil attacks in an early stage, compared to traceability. An OBU with a valid credential is not able to produce multiple messages on the same event that appear to be from different OBUs.

\paragraph{Unforgeability}
This ensures that a valid signature, including group signature and event signature, cannot be attributed to an honest member unless this member does produce it. It also guarantees that the link among multiple messages cannot be forged. Hence unforgeability of our ${\elgs}$ scheme consists of three parts: 
\begin{itemize}[leftmargin=0.15in]
\item {\em Non-frameability of the group signature} states that the adversary is unable to frame an honest OBU for producing a certain valid signature unless this OBU really did produce this signature. 
\item {\em Unforgeability of the event signature} states that the adversary is unable to forge an event signature that is related to a valid group signature produced by an honest OBU unless this OBU really did produce this signature. 
\item {\em Unforgeability of the event linkability} states that the adversary is unable to produce multiple signatures traced to the same signer on the same event, which is unlinkable, and also unable to forge the linkability with signatures signed by different OBUs. 
\end{itemize} 
The adversary defined for unforgeability is much stronger since it may fully corrupt both the opener and the issuer.

\section{Construction}\label{sec:construction}
In this section, we present the detailed construction of our {\name} protocol. The {\name} protocol includes $\setup$, 
$\ukg$, $\gsig$, $\gver$, $\esig$, $\ever$,  $\link$, $\open$ and $\judge$ algorithms, and the $(\join,\issue)$ protocol.
 %Our scheme is a hybrid scheme which combines group signatures and traditional digital signatures. The latter is called {\em event signatures} in our scheme. 
The $\setup$ algorithms initiate system parameters, and $\ukg$ creates a long-term public/private key pair for each OBU.
There are two types of signatures involved in our construction: {\em group signatures} and {\em event signatures}. 
%New algorithms $\esig$ and $\ever$ are introduced for creating and verifying event signatures.
The group signatures are created by the $\gsig$ algorithm  and verified by the $\gver$ algorithm, while event signatures are created by the $\esig$ algorithm and verified by the $\ever$ algorithm. A group signature is an anonymous signature which contains an event-linking token to allow the $\link$ algorithm to link signers. Two group signatures on the same event from the same signer contain the same event-linking token. This token is later re-used as a temporal public key, called {\em event public key}, for generating event signatures using traditional digital signatures, which do not involve the time-consuming pairing operations.   The group signature serves as an anonymous but event-linkable certificate for the event public key.  When the event is over, the event public key is automatically revoked. The $\open$ algorithm enables the opener to trace a group signature to a signer and create a tracing proof to attest to the fact. This is  for identifying and punishing a misbehaved OBU. The validity of the tracing proof can be checked by the $\judge$ algorithm. 

Our protocol is constructed using group signatures in \cite{Boneh2004} for anonymous authentication, linking token technique of Direct Anonymous Attestation \cite{Brickell:2004:DAA:1030083.1030103} for achieving event-linkability, and Schnorr signatures \cite{Schnorr:1991:ESG:2724954.2725006}
for constructing event signatures. Our scheme can be viewed as a  hybrid scheme which  combines the flexibility and convenience of a group signature with the efficiency of a traditional digital signature, which, besides stronger security and privacy and higher performance, can achieve better revocation than either of these two types of signatures alone.  
 Since we re-use the event-linking token as a public key in event signatures, the security of the combined schemes becomes non-trivial. In fact, modelling  and proving the security properties of such a protocol is challenging. 
 
% Our model for $\elgs$ scheme builds upon the model of \cite{Bellare2005,7006753}. Parties involved in our model are an authority called {\em issuer} who  authorises who can join the group and issues credentials, an authority called {\em opener}  who can revoke the signer's anonymity by opening signatures, and a group of OBUs each with a unique identity $i\in\mathbb{N}$. 

Let $(\mathbb{G}_1, \mathbb{G}_2, \mathbb{G}_T)$ be a bilinear group with prime order $p$ and $\pair: \mathbb{G}_1\times \mathbb{G}_2 \rightarrow \mathbb{G}_T$ is an efficient nondegenerate bilinear map. The protocol employs two cryptographic hash functions $\hs: \setd{0,1}^*\rightarrow \mathbb{G}_1$ and ${\hsh}:\setd{0,1}^* \rightarrow \mathbb{Z}_p$.

\subsection{Initialisation} 
The algorithm ${\setup(1^\lambda)}$ generates master secret keys for the issuer and the opener and group public parameters. 
The algorithm ${\ukg}(1^\lambda)$ produces a private/public key pair to be used as a long-term identity for an OBU. 
The algorithms proceed as follows. 
\begin{description}
\item ${\setup(1^\lambda)}$: %this algorithm creates master secret keys for the issuer and the opener, and generates public parameters. The algorithm proceeds as follows.
\begin{itemize}[leftmargin=0in]
 \item The issuer chooses $\gamma\ran\mathbb{Z}_p^*$, $g_1\ran \mathbb{G}_1\backslash\setd{1_{\mathbb{G}_1}}$, $g_2\ran \mathbb{G}_2\backslash\setd{1_{\mathbb{G}_2}}$, and computes $w = g_2^\gamma$. The issuer's master issuing key is $\mik =\gamma$, which will be used to issue membership credentials for legitimate OBUs.
 \item The opener chooses $u\ran \mathbb{G}_1\backslash\setd{1_{\mathbb{G}_1}}$, $\xi\ran\mathbb{Z}_p^*$ and computes $h = u^\xi$. The opener's master opening key is $\mok = \xi$. 
 \item The group public key is ${\gpk} = (g_1, h, u,\hs, \hsh, g_2, w)$.
\end{itemize}  
\item ${\ukg}(1^\lambda)$: An OBU $i$ chooses $y\rand\mathbb{Z}^*_p$ and sets its secret key as $\usk[i] = y$ and public key as $\upk[i] = h^y$. This key pair is an OBU's personal public key and secret key, which is used to authenticate the OBU when it registers with the issuer. The public key list $\upk$ is assumed to be public, and anyone can get an authentic public key of any user. This can be easily implemented using traditional public key infrastructure and certificates.
\end{description}

\subsection{Registration of new members}
An OBU $i$ can register with the issuer to become a legitimate group member through an interactive protocol $({\join}(\gpk, i,\upk[i],\usk[i]), \issue(\gpk,\mik, \reg, i,\upk[i]))$. Upon successful completion, the OBU $i$ becomes a group member and obtains a {\em group signing key} as its membership credential. 
    The final state of the $\issue$ algorithm is stored in the registration table at index $\reg[i]$, whereas that of the $\join$ is stored in $\gsk[i]$. The communication between the OBU and the issuer is assumed to take place over secure channels, which can be easily established using TLS/SSL.
%Assume the OBU $i$'s secret key is $y = \usk[i]$ and the corresponding  public key is $z = \upk[i]$.  
\begin{description}
\item ${\join}(\gpk, i,\upk[i],\usk[i])$:  Let  $y = \usk[i]$ and  $z = \upk[i]$.
 \begin{itemize}[leftmargin=0in]
\item OBU chooses a random $r\rand \mathbb{Z}^*_p$ and compute $c = \hsh(h,z,h^r)$ and $s = r+cy$. Send $(z, c, s)$ to the issuer. This is to prove OBU's knowledge of its secret key $y$ and show the ownership of its public key $z$.
\item Upon receiving $(x,  A)$ from the issuer, OBU checks if $\pair(A, g_2^x w) = \pair(g_1\cdot {z^{-1}}, g_2)$. If successful, OBU sets $\gsk[i] = (x, y, A)$ as its group signing key.
 \end{itemize}
 \item $\issue(\gpk,\mik, \reg, i,\upk[i])$: Upon receiving $(z, c, s)$ from the OBU, the issuer verifies whether the proof is correct by computing $\tilde c = \hsh(h, z, h^s {z}^{-c})$ and checking if $c = \tilde c$. If successful, the issuer chooses $x\rand \mathbb{Z}^*_p$ and computes $A = (g_1\cdot z^{-1})^{\frac{1}{\gamma + x}}$. The issuer stores $(x,A)$ in a registration table at index $\reg[i] = (x,  A)$ and send $(x, A)$ to the OBU.

\end{description}

\subsection{Generation of group signatures}\label{sub:gsig}
An OBU $i$ can run an algorithm $\gsig(\gpk, \gsk[i], \eid, m)$ using its group signing key $\gsk[i]$ to produce a group signature on a certain event $\eid$ with message $m$. This signature proves the knowledge of a valid group signing key in an anonymous way.
The signature also certifies an event-linking token $T$ to be used as a temporary public key in the following event signatures. 
Thus the message $m$ can be a ``hello'' message or can be simply omitted.
 Assume the OBU's group signing key is $\gsk[i] = (x, y, A)$. The algorithm proceeds as follows.
 \begin{itemize}[leftmargin=0.15in]
\item Choose $\alpha\rand \mathbb{Z}_p^*$ and compute $D = u^\alpha, B = A h^\alpha,  T = \hs(\eid)^y$. $T$ is called the {\em event-linking token}. %Set the event public key $\epk = T$.
\item Choose $r_x, r_y, r_\alpha, r_\delta \rand \mathbb{Z}_p^*$. Compute ${R}_1, {R}_2, {R}_3, {R}_4$ as follows: $R_1 = u^{r_\alpha}, R_2 = \hs(\eid)^{r_y}, R_3 = u^{r_\delta} D^{r_x}$ and $R_4 = \pair(B, g_2)^{r_x}\pair(h, w)^{r_\alpha}\pair(h, g_2)^{r_y+r_\delta}$.
\item Compute $c = \hsh(\eid, m, D, B, T, R_1, R_2, R_3, R_4)$, $s_x = cx+r_x, s_y = cy+r_y, s_\alpha = -c\alpha + r_\alpha$ and $s_\delta = -c\alpha x + r_\delta$.
%\item \red{Set an event signing key $\esk[i,\eid] = y$ and an event public key $\epk[i,\eid] = T$}
\item Output a signature $\sigma = (D, B, T, c, s_x, s_y, s_\alpha, s_\delta)$ as a group signature on the event $\eid$ and message $m$. 
\end{itemize}

The $\eid$ can be chosen by an authority, like RSU, and can also be chosen by an individual vehicle, such as the lead vehicle in a platoon or a vehicle launching a certain event such as cooperative positioning \cite{5464357,5783950}, or can be chosen by establishing an agreement among a group of vehicles. More discussions on $\eid$ can be found in \Cref{sec:app}. Each OBU can only generate an unique event-linking token $T = \hs(\eid)^y$ for an event $\eid$. Within the event $\eid$, the group signatures from the same user are linkable because they have the same event-linking token.

\subsection{Verification of group signatures}\label{sub:gver}
Any recipient of a group signature $\sigma$ on an event $\eid$ and a message $m$ can run algorithm ${\gver}(\gpk, \eid, m, \sigma)$  to check the validity of the signature. 
The algorithm proceeds as follows. Parse $\sigma$ as $(D, B, T, c, s_x, s_y, s_\alpha, s_\delta)$. Compute $\tilde{R}_1, \tilde{R}_2, \tilde{R}_3, \tilde{R}_4$ as follows: $\tilde{R}_1 = u^{s_\alpha} D^{c}, \tilde R_2 = {\hs(\eid)}^{s_y} T^{-c}, \tilde R_3 = u^{s_\delta} D^{s_x}$ and
\[\small
\tilde R_4 = \pair(B, g_2)^{s_x}\pair(h, w)^{s_\alpha}\pair(h,g_2)^{s_y+s_\delta}\bt{\frac{\pair(B, w)}{\pair(g_1, g_2)}}^c
\] Check if $c = \hsh(\eid, m, D, B, T, \tilde R_1, \tilde R_2, \tilde R_3, \tilde R_4)$. Output 1 if the check succeeds, else output 0.

\subsection{Generation of event signatures}\label{sub:esig}
After an OBU creates and broadcasts a group signature $\sigma$ with an event-linking token $T = \hs(\eid)^y$,  the OBU can use $\epk = (\hs(\eid), T)$ as an {\em event public key} to produce event signatures that can be verified against this temporary public key.  The event signatures are generated using algorithm $\esig(\usk[i],\eid, \epk, m_e)$, which proceeds as follows. Let $y = \usk[i]$ be the $i$-th OBU's secret key. Choose a random $r\rand \mathbb{Z}^*_p$. Compute $R = \hs(\eid)^r$ and $c_e = \hsh(\eid, m_e, \epk, R)$. Output an event signature $\sigma_e = (s_e,c_e)$ where $s_e = r +  y c_e$.

 The event signatures are the traditional digital signatures which do not involve pairing operations and, for this reason, are much more efficient compared to group signatures. An OBU can use event signatures to, for example, update and sign its status information (e.g., GPS location, speed). Note that the event public key is self-certified in the group signature and is bound with the event $\eid$. When the event is over, the event public key is automatically revoked and is no longer valid.
%It is simple and natural to re-use the event-linking token as a temporary public key, but the security modelling and proofs are actually challenging. We give the full proofs in the appendix.  

\subsection{Verification of event sigantures}\label{sub:ever}
Any recipient of an event signature $\sigma_e$ can run the verification algorithm
 $\ever(\eid,\epk, m_e,\sigma_e)$  to check the validity of the signature using the corresponding event-linking token. The algorithm  proceeds as follows. Parse $\sigma_e = (s_e,c_e)$. Compute $\tilde R = \hs(\eid)^{s_e} \epk^{-c_e}$ and $\tilde c = \hsh(\eid, m_e, \epk, R)$. If $c_e = \tilde c$, then output 1, else output 0.

\subsection{The event-linking algorithm on group signatures}\label{sub:link}
Any recipient of two group signatures $\sigma_0, \sigma_1$ on an event $\eid$ can run a linking algorithm $\link(\eid, m_0, \sigma_0, m_1,\sigma_1)$ to check if the two signatures are signed by the same OBU. 
The algorithm proceeds as follows. Parse $\sigma_b$ as $(D_b, B_b, T_b, c_b, s_{x,b}, s_{y,b}, s_{\alpha,b}, s_{\delta,b})$ for $b = 0,1$. The algorithm compares the event-linking tokens $T_0, T_1$: if $T_0 = T_1$, then output 1 else, output 0.

An OBU $i$ can  generate at most one event-linking token for a certain event. This guarantees that all the messages signed by the OBU using  group signatures and event signatures on a certain event are linkable by any recipient, which effectively thwarts Sybil attacks as early as possible. However, these group signatures and event signatures do not leak any information about the OBU's long-term identity $\upk[i]$ to the public, which preserves the OBU's anonymity.     

\subsection{Opening group signatures}\label{sub:trace}
The opener can run an opening algorithm ${\open}(\gpk, \mok, \reg, \sigma)$ to recover the identity of the signer of a group signature $\sigma$ and produce a tracing proof $\pi$ attesting to this fact. The algorithm proceeds as follows. 
Parse $\sigma$ as $(D,B,T,c,s_x, s_y,s_\alpha,s_\delta)$. Computes $A = B D^{-\xi}$ and finds $i$ such that $\reg[i] = (x, A)$ for some $x$.  $i$ is the identity of the original signer. Then the algorithm produces proof that this signature is indeed produced by $i$.
Choose $r\rand \mathbb{Z}^*_p$, compute the proof as $\pi = (K,s, c, x)$ where $K = D^\xi,  c = \hsh(\sigma,K,u^r, D^r), s = r+\xi c$, and return $(i,\pi)$. % If the algorithm failed to trace the signature to a particular group member, then returns $(\bot,\bot)$.

\subsection{Judging tracing proofs}
Any recipient of a group signature $\sigma$ and a tracing proof $\pi$ can run a judging algorithm  ${\judge}(\gpk,i,\upk[i],\sigma,\pi)$ to verify if $\pi$ is a valid proof that OBU $i$ produced $\sigma$. The algorithm proceeds as follows. Parse $\sigma$ as $(D,B,T,c,s_x, s_y, s_\alpha, s_\delta)$ and $\pi = (K,s, c, x)$.
Let $z = \upk[i]$.  Compute $\tilde c = \hsh(\sigma, K, u^s h^{-c}, D^s K^{-c})$. If $\tilde c = c$ and $\pair(B K^{-1}, w g_2^x) = \pair(g_1 z^{-1}, g_2)$ output 1, else output 0.

\section{Applications in vehicular networks}\label{sec:app}
In this section, we describe how to apply the proposed {\name} protocol to exemplar applications of  V2X communications.  %Recall that vehicles can communicate with each other and with roadside unit (RSU) through wireless connections such as DSRC. 
%Recall that our scheme enables a vehicle to generate a group signature on a certain event to self-certify an event-linking token as a temporary public key, and then to sign its status information with event signatures which are verified by using the temporary public key.
The discussion here mainly focuses on how to instantiate events in each application which decides the way the vehicle messages are linked. Generally speaking, the realisation of events is flexible and can vary from case to case. The events have to be updated periodically to protect a vehicle's long-term privacy. The more frequently the events are changed, the stronger privacy and weaker linkability. The events can be chosen by an authority, like RSU, and can also be chosen by an individual vehicle, such as the lead vehicle in a platoon or a vehicle launching a certain event such as cooperative positioning \cite{5464357,5783950}, or can be chosen by establishing an agreement among a group of vehicles.
To demonstrate how the proposed {\name} scheme can be used  in V2X system applications, we consider two exemplar  use cases in vehicular networks: intersection management \cite{6728435,7313162} and cooperative awareness messages \cite{etsi2014}.

\subsection{Intersection management}
When an authority RSU is nearby, events can be created by the authority.  We consider the intersection management \cite{6728435,7313162} for scheduling connected and autonomous vehicles to cross a traffic intersection safely. These techniques utilise the communication between a central controller and vehicles, while the conventional traffic control methods, such as stop signs and traffic lights, have been removed.

An intersection controller is a program that runs on a centralised infrastructure physically located at the intersection. The controller only supervises vehicles located within a certain distance of the intersection.
A centralised intersection scheduling algorithm \cite{6728435,7313162} proceeds as follows:
\begin{enumerate}[leftmargin=0.2in]
\item When a vehicle approaches an intersection area, the vehicle periodically broadcasts its status information, such as the vehicle's position, heading, desired velocity,  the brake or accelerator input, vehicle profile and future path.
\item If the current states of the controlled vehicles lead to an inevitable collision, the controller broadcasts a safe input to override the inputs of the controlled vehicles. Though the main goal of such intersection scheduling algorithms is to avoid collisions, the controller can also consider some mechanisms to optimise the system performance, such as reducing the aggregate fuel consumption. % by slowing down a light vehicle instead of a bus or a heavy truck. 
\item The controller samples the next set of vehicles and repeats the above steps.
\end{enumerate}

In this application scenario, events can be generated and updated by the controller. %As shown in \Cref{fig:overview}, 
The controller can construct an event  as $\eid:=\mbox{``intersection location || timestamp''}$ to avoid using the same $\eid$ at different intersections. This prevents the vehicles from being tracked at different intersections. The controller broadcasts the current $\eid$.  The authenticity of the $\eid$ from RSU can be easily achieved using digital signatures and PKI certificates since RSU does not need privacy protection.  When a vehicle $i$ approaches the intersection area, it creates and broadcasts a group signature $\sigma_i$ on $\eid$. The event-linking token $T_i$ in $\sigma_i$ serves as a temporary identifier of the vehicle $i$ for the scheduling purpose. 
The vehicle continues to periodically update and sign its status information %(e.g., position, heading, velocity, path, vehicle type) 
using the event signatures. The controller computes a safe input $u_i$ for vehicle $i$  based on the control strategy mentioned in the above step 2).  Then the controller signs and broadcasts $(T_i,u_i)$ to inform the vehicle $i$ to change its control input to $u_i$. At the sender side, the group signature $\sigma_i$ is computed only once for each event. In case of packet loss, $\sigma_i$ can be re-broadcast once every $n$ messages to enhance robustness \cite{Calandriello:2007:ERP:1287748.1287752}. At the receiver's side, $\sigma_i$ is verified upon the first reception and stored for validating the following event signatures. 

%Suppose at an intersection $A$ there is an event $\eid$ for scheduling vehicles to safely cross the intersection at a time $t$. When a vehicle $V$ approaches the intersection $A$, it sends a group signature which contains an event linking token on $\eid$. Then the vehicle updates its status information (i.e., position, heading, speed, acceleration, etc.) and signs status messages with a traditional digital signature scheme. The intersection scheduler computes an optimised crossing order for all the vehicles in control.  These signatures are verified against the event-linking token which behaves like an ephemeral public key. The movement of the vehicle is trackable for the intersection scheduler and other nearby vehicles which help avoid collisions. When the same vehicle later participants in an scheduling protocol at an intersection $B$, a new event will be used and the movement of this vehicle becomes unlinkable with its movement at the intersection $A$, which preserves the vehicle's long-term privacy.

Different intersections use different events to ensure that a vehicle won't be identified among intersections. But a vehicle's movement can be tracked within the intersection region in a short time period to enable the controller to run the scheduling algorithms and to monitor that the vehicle is following the controller's output. The controller can decide when to change to a new event according to the traffic densities. If there is a traffic jam in the intersection area, for example, when the average waiting time of a vehicle is about an hour, then the event will be updated less frequently and may last for hours. When the traffic density is low, events can be changed more frequently. The controller can choose to change events at the point when its scheduling algorithms start to sample a new set of vehicles, i.e., step 3) in the above description. If the controller uses an event for too long, the multiple activities of the same vehicle passing through the intersection can be identified. For example, a vehicle crossing the intersection at 9:00 in the morning and later passing through this intersection again at 17:00 in the afternoon if the controller uses the same event for a day.

\subsection{Cooperative awareness messages}
This application enables vehicles to improve their awareness of the key road traffic events by exchanging status information, e.g., position, dynamics and attributes, in the periodically transmitted cooperative awareness messages (CAM) \cite{etsi2014}.
The information in CAMs forms the basis for many safety applications in V2X communications, such as collision avoidance, traffic condition warning, and hazard warning \cite{etsi2009}.
According to the European standard \cite{etsi2014}, the CAM generation interval is between $0.1s \sim 1s$, and the data processing time of CAM should not exceed 50ms in order to ensure the effectiveness of the safety messages.

Note that all the vehicles nearby have to use the same $\eid$ within a certain time period to track a vehicle's local movements to avoid collisions. The $\eid$ should also be updated periodically to protect a vehicle's long-term privacy. A simple way to deal with these problems is to pre-define $\eid$ as a timestamp which is valid in a certain timeslot. For example, from 10:00 am to 10:10 am on 1 March 2017, all the vehicles will use the $\eid:= ``201703011000"$, and then switch to a new $\eid:=``201703011010"$ from 10:10 am to 10:20 am, and so on. This approach for generating $\eid$ does not rely on the presence of any central authority such as RSU. Instead, all the vehicles agree in advance to produce and update their $\eid$s using a pre-defined method. 
The advantage of this method is all the $\eid$ that will be used in future become predictable. As a result, the time-consuming group signatures in our scheme can be computed offline when a vehicle does not have heavy computation work,  e.g., parked at the garage. If the $\eid$ is changed every 10 mins, then a total number of 144 $\eid$s are needed for a whole day's use. Using the evaluation results in next \Cref{sec:implementation}, computing 144 group signatures for these $\eid$s will only take  1.83s on the laptop and 22.8s on Raspberry Pi 3. Thus the offline computation of the group signatures can be made without being noticeable to the users.

%Begin [yy-0213] Add new section: "Implementation".
%Written by Yan Yan (y.yan@bristol.ac.uk)

\section{Implementation}\label{sec:implementation}
We have implemented the proposed $\elgs$ scheme using the Pairing Based Cryptography (PBC) Library\cite{lynn2007implementation}. Our tests use a {\em d-type} curves ($d224$) from PBC library. %The source code is available at \cite{eggs}.
Although currently there is no agreement about a vehicle's on-board hardware capabilities, % to reduce production cost,  %the vehicle's computing power is restricted by the fact that it is powered by the fuel and battery, and thus does not have a stable power source. 
we present illustrative measures taken from an experiment done with an ASUS ZenBook3 UX390UA\footnote{Powered by Intel i7500U with 8GB RAM} and  a Raspberry Pi 3 (model B)\cite{RP3}\footnote{Powered by a 1.2GHz 64-bit quad-core ARMv8 CPU and 1GB RAM.}.

%\subsection{Signature Size}
%\Cref{Tbl: NumOfEle} summarises the number of elements in each field. 

\begin{table}
	\centering
	\resizebox{\columnwidth}{!}{
	\begin{tabular}{c|c|c|c|c}
		\hline
	 & BBS \cite{Boneh2004} & TAA \cite{5719272}  & PS-OL & Ours \\
		\hline
	 Group Sig. & $3\mathbb{G}_1 + 6\mathbb{Z}_p$ & 		     \begin{tabular}{c}
	 $5\mathbb{G}_1 + 3\mathbb{Z}_p$ (v1)\\
	 $7\mathbb{G}_1 + 3\mathbb{Z}_p$ (v2) \\
	 \end{tabular} & $3\mathbb{G}_1 + 5\mathbb{Z}_p$ & $3\mathbb{G}_1 + 5\mathbb{Z}_p$ \\
	 \hline
	 Event Sig. & \multicolumn{3}{c|}{N/A} & $2\mathbb{Z}_p$ \\
		\hline
	\end{tabular}
	}
	\caption{Comparison of signature length. \label{Tbl: SigLen}}
\end{table}

In our $\elgs$ scheme, the group signatures consists of 3 elements from $\mathbb{G}_1$ and 5 elements from $\mathbb{Z}_p$ which retains the normal size of a short group signature scheme \cite{Boneh2004,5719272,7006753} as shown in \Cref{Tbl: SigLen}. The event signatures are significantly shorter, consisting of 2 elements from $\mathbb{Z}_p$. The actual size when implemented on curve d224 is summarised in \Cref{Tbl: EgsSigSize}.
The compressed size is obtained using PBC compression algorithm on the elements in $\mathbb{G}_1$.
\Cref{Tbl: BasePerformance} shows the running time of group operations including multiplication, exponentiation and pairing.

%\begin{table}
%	\centering
%	\begin{tabular}{c|c|c|c|c}
%		\hline
%		& $\mathbb{G}_1$ & $\mathbb{G}_2$ & $\mathbb{G}_T$ & $\mathbb{Z}^*_p$ \\
%		\hline
%		Group signatures & 3 & 0 & 0 & 5 \\ 
%		\hline
%		Event signaures & 0 & 0 & 0 & 2 \\
%		\hline
%%		PS-OL Signature\cite{hwang2013short} & 3 & 0 & 0 & 5 \\
%%		\hline
%	\end{tabular}
%	\caption{Number of elements of our $\elgs$ scheme}\label{Tbl: NumOfEle}
%\end{table}

%\begin{table}
%	\centering
%	\begin{tabular}{c|c|c|c}
%		\hline
%		& & Full & Compressed \\
%		\hline
%		\multirow{2}{*}{d159} & Group signature & 220 & 163 \\
%		 & Event signature &  40 &  40 \\
%		\hline
%		\multirow{2}{*}{d224} & Group signature & 308 & 227 \\
%		 & Event signature & 56 & 56 \\
%		\hline
%	\end{tabular}
%	\caption{$\elgs$ signature size (bytes) \label{Tbl: EgsSigSize}}
%\end{table}

\begin{table}
	\centering
	\begin{tabular}{c|c|c}
		\hline
	 & Full & Compressed \\
		\hline
	 Group Sig. & 308 & 227 \\
	 Event Sig. & 56 & 56 \\
		\hline
	\end{tabular}
	\caption{Size of signatures in $\elgs$ (bytes). \label{Tbl: EgsSigSize}}
\end{table}

%\subsection{Performance evaluation}

{\em Optimisation of $\gsig$:} This algorithm can be optimised to be pairing-free. Specifically, notice that the $R_4$ can be transformed to:
\[\small
	\begin{split}
		R_4 &= \pair(B, g_2)^{r_x} \pair(h, \omega)^{r_\alpha} \pair(h,g_2)^{r_y + r_\delta} \\
		    &= \pair(A, g_2)^{r_x} \pair(h, \omega)^{r_\alpha} \pair(h,g_2)^{\alpha \cdot r_x + r_y + r_\delta}
	\end{split}
	\]
The pairings $\pair(A, g_2)$, $\pair(h, \omega)$ and $\pair(h, g_2)$ in $R_4$ are reusable and does not depend on any variable generated during the procedure; therefore they can be computed in advance to reduce computation.

{\em Optimisation $\gver$:} The number of pairing operations can be reduced by modifying $\tilde R_4$ as below:
\[
\small
\begin{split}
\tilde R_4 & = \pair(B, g_2)^{s_x}\pair(h, w)^{s_\alpha}\pair(h, g_2)^{s_y + s_\delta} \bt{\frac{\pair(B, w)}{\pair(g_1, g_2)}}^c \\
& = \pair(\frac{B^{s_x} h^{s_y+s_\delta}}{g_1^c}, g_2)~ \pair(h^{s_\alpha} B^c, w)
\end{split}
\]
\Cref{Tbl: NumOfGO} summarised the number of group operations of $\gsig$ and $\gver$ after optimisation. Here $\mul_{\mathbb{G}_1}$, $\mul_{\mathbb{G}_T}$, $\iexp_{\mathbb{G}_1}$ and $\iexp_{\mathbb{G}_T}$ are multiplications on $\mathbb{G}_1$ and $\mathbb{G}_T$, and exponentiations on $\mathbb{G}_1$ and $\mathbb{G}_T$, respectively.

Despite the fact that processors we tested are multi-cored, neither PBC nor our implementation has utilised any parallelisation; hence it would be reasonable to expect our profiles being applicable even for single core processors. Our implementation does not have any curve dependency; therefore it can be easily ported to any other qualified curves by simply specifying a different curve parameter file.

\begin{table}[!htbp]
	\centering
%	\tablinesep=2ex\tabcolsep=10pt
	\begin{tabular}{c|c|c|c}
		\hline
		& & ZenBook  & Raspberry Pi 3 \\
		\hline
		\multirow{3}{*}{\mul} & ${\mathbb{G}_1}$  & 0.003 &  0.02 \\
		& ${\mathbb{G}_2}$  & 0.02 &  0.23 \\
		& ${\mathbb{G}_T}$  & 0.005 &  0.07 \\
		\hline
		\multirow{3}{*}{\iexp} & ${\mathbb{G}_1}$  & 0.92 &  5.65 \\
		& ${\mathbb{G}_2}$  & 6.48 & 60.47 \\
		& ${\mathbb{G}_T}$  & 2.35 & 26.52 \\
		\hline
		{\it pairing} &            & 6.19 & 61.93 \\
		\hline
	\end{tabular}
	\caption{Performance of group operations (ms). \label{Tbl: BasePerformance}}
\end{table}

\begin{table}[!htbp]
	\centering
	\begin{tabular}{c|c}
		\hline
		$\gsig$		& $3\mul_{\mathbb{G}_1} + 4\iexp_{\mathbb{G}_1} + 2\mul_{\mathbb{G}_T} + 3\iexp_{\mathbb{G}_T} $  \\
		\hline
		$\gver$	& $6\mul_{\mathbb{G}_1} + 11\iexp_{\mathbb{G}_1} + 1 \mul_{\mathbb{G}_T} + 2 {\it pairing} $  \\
		\hline
		$\esig$		& $1 \iexp_{\mathbb{G}_1}$\\
		\hline
		$\ever$	& $1 \mul_{\mathbb{G}_1} + 2 \iexp_{\mathbb{G}_1}$  \\
		\hline
	\end{tabular}
	\caption{Numbers of group operations of signatures in $\elgs$.\label{Tbl: NumOfGO}}
\end{table}

\Cref{Tbl: EggsPerformance} presents the timing profile of our $\elgs$ protocol.
%We also profiled ECDSA of OpenSSL implementation using curve secp192k1 on the same devices as a comparison.
Each result is calculated on 1000 samples. $\esig$ and $\ever$ are 10 times faster than $\gsig$ and $\gver$ respectively. $\gsig$ and $\gver$ only needs to be performed once for each event, while $\esig$ and $\ever$ are for the frequent use of  safety message authentication. Note that the time for $\open$ only includes the time for recovering $A$; the search for user identity is the equality comparison which does not involve any cryptographic operations and can be optimised to be $O(1)$ using $A$ as index in the database.
%The $\link$ algorithm is negligible on Zenbook3 ($< 0.001\text{ms}$) and takes only $0.004$ms on Raspberry Pi 3.

%Despite the fact that our E-signatures are essentially performing the same operations as ECDSA signatures. therefore it would be a space left to further optimise the implementation. However, in order to maintain the flexibility of choosing different curves, we leave this optimisation as a future task.

\begin{table}[!htbp]
	\centering
		\begin{tabular}{|c|c|c|c|c|c|}
		\hline
		\multicolumn{6}{|c|}{ZenBook3} \\
		\hline
		 $\gsig$  & $\gver$ & $\esig$ & $\ever$  & $\link$ & $\open$ \\
		 12.72 & 26.55 	&  1.25 &  2.31 & <0.001 & 0.001 \\
		\hline
%		ECDSA		&   0.26 &   0.32 \\
%		\hline
	\end{tabular}
	\\
\vspace{0.5cm}
	\begin{tabular}{|c|c|c|c|c|c|}
		\hline
	\multicolumn{6}{|c|}{Raspberry Pi 3}  \\
		\hline
		$\gsig$ & $\gver$  & $\esig$  & $\ever$  &  $\link$  & $\open$ \\
		158.62 & 201.84 &  16.02 &  24.66 & 0.004 & 0.012 \\
		\hline
%		ECDSA		&   2.58 &   2.85 \\
%		\hline
	\end{tabular}
	\caption{Timing profile of $\elgs$ (ms). \label{Tbl: EggsPerformance}}
\end{table}

%End [yy-0213]

\section{Security analysis}\label{sec:secanalysis}
In this section, we informally describe the definitions of the security requirements of anonymity, traceability, event linkability and unforgeability and analyse that our protocol satisfies the correctness and these security requirements.
Each security property is defined as an experiment conducted between an adversary and a challenger (see \Cref{fig:games} in the appendix). % \Cref{sec:securityproofs}
The adversary can win the experiment when certain conditions are satisfied.  The protocol satisfies a security property defined by an experiment when any adversary's winning probability is negligible.
For the sake of readability, the detailed security models and proofs can be found in the appendix.

\begin{theorem}\label{thm:correct}
The {\name} protocol is correct.
\end{theorem}

We show the correctness of the proposed protocol.
First, the signing and verifying algorithms are correct because of the following relations:
\[
\small
\begin{split}
R_1 & = u^{r_\alpha} = u^{s_\alpha + c\alpha} = u^{s_\alpha} D^c \\
R_2 & = \hsh(\eid)^{r_y} = \hsh(\eid)^{s_y-cy} = \hsh(\eid)^{s_y} T^{-c} \\
R_3 & = u^{r_{\delta}} D^{r_x} = u^{s_\delta + c\alpha x} D^{s_x-cx} = u^{s_\delta} D^{s_x} \\
R_4 & =\pair(B, g_2)^{r_x}\pair(h, w)^{r_\alpha}\pair(h, g_2)^{r_y+r_\delta} \\
& = \pair(B, g_2)^{s_x-cx}\pair(h, w)^{s_\alpha+c\alpha}\pair(h, g_2)^{s_y-cy + s_\delta + c\alpha x} \\
& = \pair(B, g_2)^{s_x}\pair(h, w)^{s_\alpha}\pair(h, g_2)^{s_y + s_\delta} \bt{\frac{\pair(B, w)}{\pair(g_1, g_2)}}^c
\end{split}
\]
The last equation is because $\pair(g_1,g_2) = \pair(B,w)\pair(B,g_2)^x \\ \pair(h,w)^{-\alpha} \pair(h,g_2)^{y-\alpha x}$.
The $\open$ algorithm is correct because of $v^\alpha = u^{\xi\alpha}$.
The $\judge$ algorithm is correct because $\pair(A,w g_2^x) = \pair(g_1 h^{-y}, g_2)$ when $A = (g_1 h^{-y})^{\frac{1}{\gamma+x}}$.
The $\esig$ and $\ever$ are correct because $R = \hs({\eid})^r = \hs(\eid)^{s_e-c_ey} = \hs(\eid)^{s_e} T^{-c}$.
The $\link$ algorithm is correct because, it outputs 1 when $i_0 = i_1$ since $T_0 = T_1 = \hs(\eid)^y$. When $i_0,\neq i_1$, the probability that $y_{i_0} = y_{i_1}$ is negligible since $y_{i_0}$ and $y_{i_1}$ are chosen uniformly at random. Then we know that $T_0 = \hs(\eid)^{y_{i_0}} \neq T_1 = \hs(\eid)^{y_{i_1}}$ with overwhelming probability when $i_0,\neq i_1$. Therefore, the probability that $\link$ outputs 1 when $i_0 \neq i_1$ is negligible.

\begin{theorem}\label{thm:anony}
The {\name} protocol is anonymous under the XDH assumption.
\end{theorem}
Following \cite{Bellare2003}, anonymity is defined in a way that the adversary does not need to recover the identity of a signer but only distinguishes which of two signers of its choice signed a target message of its choice. This definition covers both identity anonymity and unlinkability. In the experiment of defining anonymity, %an adversary chooses two group members $i_0, i_1$, an event $\eid$ and a message $m$, obtains from the challenger a group signature signed on the message $m$ and event $\eid$ with either member's group signing key. 
the adversary can learn the master issuing key, and the personal secret and group signing keys of any users except the challenge members. 
%The adversary wins the experiment if guesses correctly which member the challenge has chosen to sign. The protocol satisfies the anonymity when any adversary's wining probability is negligible.
The adversary is not given the master opening key, otherwise identifying the signer would become trivial. 
In the proof, we construct an adversary who breaks the XDH assumption from an adversary who breaks the anonymity of the protocol.
Intuitively, in a group signature $\sigma = (D,B,T,c,s_x, s_y,s_\alpha, s_\delta)$, the part $(D,B) = (u^\alpha, A\cdot h^{\alpha})$ is an encryption of the signer's identity and the other parts can be simulated using randoms. Since the encryption does not leak any information about $A$, the adversary cannot use it to link two signatures of the same signer or different signers. This justifies why the protocol satisfies anonymity. 

\begin{theorem}\label{theorem:trace}
The {\name}  protocol is traceable under the $q$-SDH assumption.
\end{theorem}

In the experiment of defining traceability, an adversary generates a group signature $\sigma$ on an event $\eid$ and a message $m$ of his choice. The adversary can learn the master 
opening key as well as any user's secret key and group signing key. 
The adversary wins if $\sigma$ opens to an invalid member  or the tracing proof does not verify. 
Note that the adversary is not given the master issuing key since this would allow the adversary to create dummy group members which cannot  be tracked. 
The proof goes by constructing an adversary $\B$ which breaks the $q$-SDH assumption using an adversary $\A$ which breaks traceability as a subroutine. If the signature produced by the adversary $\A$ is valid, then $\B$ can extract a group signing key $(\Delta x, \Delta y, A = (g_1 h^{-\Delta y})^{\frac{1}{(\gamma + \Delta x)}})$ and event-inking tokens $T = \hs(\eid)^{\Delta y}$  using forking lemma. If this group signing key does not belong to any valid group member, then $\B$ can use to construct a solution for the $q$-SDH problem.

 \begin{theorem}\label{thm:eventlink}
The {\name} protocol is event linkable under the $q$-SDH assumption.
\end{theorem}

In the experiment of defining  event-linkability, an adversary can learn the master opening key and any user's secret key and group signing key.  The adversary produces two group signatures on the same event using group signing keys of his choice. The adversary wins if the two signatures are opened to the same signer but cannot be lined by the {\link} algorithm, or they are opened to different signers but are linked by the {\link} algorithm. 
The proof goes by constructing an adversary $\B$, which breaks the $q$-SDH assumption using an adversary $\A$, which breaks the event-linkability as a subroutine. If the two signatures produced by the adversary $\A$ is valid, then $\B$ can exact two group signing keys $(\Delta x_b, \Delta y_b, A_b = (g_1 h^{-\Delta y_b})^{\frac{1}{(\gamma + \Delta x_b)}})$ and event-inking tokens $T_b = \hs(\eid)^{\Delta y_b}$ for $b=0,1$  using forking lemma. If the two group signing keys are the same, then  $T_0 = T_1$ and the $\link$ algorithm should output 1. Otherwise, $T_0 \neq T_1$ and the $\link$ algorithm should output 0. %This contradicts to the assumption that $\A$ breaks the event linkability.

 \begin{theorem}\label{thm:gsignonframe}
The {\name} protocol is non-frameable under the DL assumption.
\end{theorem}
In the experiment of defining non-frameability of group signatures, an adversary can compromise the master issuing key, the master opening key and any user's secret key and group signing key. The adversary produces a group signature $\sigma$ on an event $\eid$ and a message $m$ of his choice and a proof $\pi$. The adversary wins if the signature and the proof are both valid and the signature is traced to a member whose group signing key is not compromised.  The proof of this theorem goes by constructing an adversary $\B$, which breaks the DL assumption using an adversary $\A$, which breaks the non-frameability. %Similarly, if the signature produced by the adversary $\A$ is valid, then $\B$ can extract a group signing key $(\Delta x, \Delta y, A = (g_1 h^{-\Delta y})^{\frac{1}{(\gamma + \Delta x)}})$.

\begin{theorem}\label{thm:eventunforge}
The {\name} protocol is event-unforgeable under the DL assumption.
\end{theorem}

In the experiment for defining the unforgeability of the event signature, an adversary can compromise the master issuing key, the master opening key, and any user's secret key and group signing key. The adversary produces an event signature $\sigma_e$, which is related to a valid group signature $\sigma$ generated by a member $k$. The adversary wins if the event signature is valid and the member $k$'s group signing key is  not compromised by the adversary. The proof of this theorem goes by constructing an adversary $\B$, which breaks the DL assumption using an adversary $\A$, which breaks the unforgeability. %Similarly, if the signature produced by the adversary $\A$ is valid, then $\B$ can extract a group signing key $(\Delta x, \Delta y, A = (g_1 h^{-\Delta y})^{\frac{1}{(\gamma + \Delta x)}})$.

 \begin{theorem}\label{thm:linkunforge}
The {\name} protocol is link-unforgeable under the DL assumption.
\end{theorem}
In the experiment of defining the unforgeability of the event linkability, an adversary can compromise the master issuing key, the master opening key, and any user's secret key and group signing key. The adversary produces two group signatures $\sigma_b$ and two proofs $\pi_b$ for $b=0,1$. The adversary wins if either the two signatures are traced back to two different signers, but they are linked by {\link}, or the two signatures are traced back to a single signer but {\link}  failed to link them.
Note that the experiment defining the unforgeability of the event linkability might look similar to the one for event linkability in Theorem \ref{thm:eventlink}. But the former definition considers a stronger adversary who can compromise the master issuing key and the master opening key and has access to oracles for producing group signatures and event signatures and for modifying the issuer's registration table.

\section{Conclusion}\label{sec:conclusion}
%To enhance security, privacy and efficiency for securing applications in V2X communications, we propose a hybrid group signature scheme with event-based linkability and revocation.  Our hybrid scheme combines the flexibility and convenience of a group signature with the efficiency of a traditional digital signature. %We formalise security properties of our scheme as anonymity, traceability, event linkability and unforgeability in a model for dynamic group signatures, and prove these properties in the random oracle model.
We propose an efficient anonymous protocol which supports event-based linkability and revocation. Our protocol introduces a new method for generating temporary public keys to reduce the computational complexity of cryptographic operations.
Our concept of event linkability generalises the previous notion of short-term linkability \cite{Golle:2004:DCM:1023875.1023881} by enabling flexibility on how the vehicles are linked in different situations.
We illustrate how to apply our protocol in vehicular communications by  two exemplar use cases. % In particular, we propose a pre-computation mechanisms for generating time-consuming group signatures in advance when the future events are predictable.

\section*{Acknowledgment}
This work has been funded by the UK EPSRC as part of the PETRAS IoT Research Hub - Cybersecurity of the Internet of Things grant no: EP/N02334X/1.

\bibliographystyle{plain}
\bibliography{ref}

\appendices

\section{Security model and analysis}\label{sec:correctsecurity}
The security requirements (correctness, anonymity, traceability, event linkability and unforgeability) are formulated 
via experiments described in \Cref{fig:games} where an adversary has access to certain oracles defined in \Cref{fig:oracles}.

  %The formal definitions of the oracles are given in Figure \ref{fig:oracles}.
%Note that there is no oracle for the $\link$ algorithm because $\link$ does not use any secret and can be called by the adversary itself.

\begin{figure*}[pt]
\scriptsize
%\hspace*{-10mm}
\begin{minipage}[t]{.45\textwidth}
%\begin{multicols}{2}
$\addu(i)$:
\begin{itemize}[leftmargin=0.2in,topsep=0pt]
\item If $i\in \hu$ then return $\bot$
\item $\hu=\hu\cup\setd{i}$
\item $dec^i = {\sf cont}$; $\gsk[i] = \bot$
\item $(\upk[i],\usk[i])\rand \ukg(1^\lambda)$
\item $St_{jn}^i = (\gpk, \upk[i],\usk[i])$
\item $St_{iss}^i = (\gpk, \mik, \upk[i])$; $M_{jn} = \bot$
\item $(St_{jn}^i, M_{iss}, dec^i)\rand \join(St_{jn}^i, M_{jn})$
\item While $dec^i = {\sf cont}$ do
\begin{itemize}[leftmargin=0.1in]
\item[] $(St_{iss}^i, M_{jn}, dec^i) \rand \issue(St_{iss}^i, M_{iss}, dec^i)$
\item[] If $dec^i = {\sf accept}$ then $\reg[i]\rand St_{iss}^i$
\item[] $(St_{jn}^i, M_{iss}, dec^i)\rand \join(St_{jn}^i, M_{jn})$
\end{itemize}
\item Endwhile
\item $\gsk[i] = St_{jn}^i$
\item Return $\upk[i]$
\end{itemize}

\noindent\makebox[\linewidth]{\rule{\textwidth}{0.3pt}}

$\ogsk(i)$
\begin{itemize}[leftmargin=0.2in,topsep=0pt]
\item Return $\bot$ if $i\notin \hu\backslash \bu$ or $(i,*,*,*)\in \cl$ \footnote{\scriptsize We write $(i,*,*,*)\in\cl$ for ``$(i,\eid,m,\sigma)\notin\cl$ for some $\eid,m,\sigma$''. And similarly $(i,\eid,*,*)\in \cl$ short for ``$(i,\eid,m,\sigma)\in\cl$ for some $m,\sigma$''.}
\item $\bu := \bu\cup\setd{i}$
\item Return $(\gsk[i],\usk[i])$
\end{itemize}

\noindent\makebox[\linewidth]{\rule{\textwidth}{0.3pt}}

$\gsig(i,\eid,m)$
\begin{itemize}[leftmargin=0.2in,topsep=0pt]
\item If $i\notin \hu$ then return $\bot$
\item If $\gsk[i]=\bot$ then return $\bot$
\item If $(i,\eid,*,*)\in\cl$ then return $\bot$
\item $\sigma\rand \gsig(\gpk,\gsk[i],m)$
\item $\isl[\sn] := (i,\eid,m,\sigma)$
\item $\sn:=\sn+1$
\item Return $\sigma$
\end{itemize}

\noindent\makebox[\linewidth]{\rule{\textwidth}{0.3pt}}

$\rreg(i)$:
\begin{itemize}[leftmargin=0.2in,topsep=0pt]
\item Return $\reg[i]$
\end{itemize}

\end{minipage}%
~~~~\vline~~~~
\begin{minipage}[t]{.45\textwidth}

$\wreg(i,val)$:
\begin{itemize}[leftmargin=0.2in,topsep=0pt]
\item $\reg[i] := val$
\end{itemize}

\noindent\makebox[\linewidth]{\rule{\textwidth}{0.3pt}}

$\esig(k, m_e)$
\begin{itemize}[leftmargin=0.2in,topsep=0pt]
\item If $k\geq 1$,
\begin{itemize}[leftmargin=0.1in]
\item[] Return $\bot$ if $\isl[k]=\bot$
\item[] Assume $\isl[k] = (i,\eid,m,\sigma)$
\item[] Parse $\sigma$ to extract the event public key $\epk$
\item[] $\sigma_e\rand \esig(\usk[i],\eid, \epk, m_e)$
\item[] Return $\sigma_e$
\end{itemize}
\item If $k=0$,
\begin{itemize}[leftmargin=0.1in]
\item[] Return $\bot$ if $\cl=\emptyset$
\item[] Assume $(i,\eid,m,\sigma)\in \cl$ with $\sigma\neq \bot$
\item[] Parse $\sigma$ to extract the event public key $\epk$
\item[] $\sigma_e\rand \esig(\usk[i],\eid, \epk,m_e)$
\item[] Return $\sigma_e$
\end{itemize}
\item Return $\bot$
\end{itemize}

\noindent\makebox[\linewidth]{\rule{\textwidth}{0.3pt}}

$\sndu(i, M_{in})$
\begin{itemize}[leftmargin=0.2in,topsep=0pt]
\item If $i\notin \hu$
\begin{itemize}[leftmargin=0.1in]
\item[] $\hu := \hu \cup \setd{i}$
\item[] $(\upk[i],\usk[i])\rand \ukg(1^\lambda)$
\item[] $\gsk[i]:=\bot; M_{in}:=\bot$
\item[] $St_{jn}^i := (\gpk,\upk[i],\usk[i])$
\end{itemize}
\item $(St_{jn}^i, M_{out}, dec)\rand \join(St_{jn}^i,M_{in})$
\item If $dec = {\sf accept}$ then $\gsk[i] := St_{jn}^i$
\item Return $(M_{out}, dec)$
\end{itemize}

\noindent\makebox[\linewidth]{\rule{\textwidth}{0.3pt}}

$\ch(i_0, i_1, \eid, m)$
\begin{itemize}[leftmargin=0.2in,topsep=0pt]
\item Return $\bot$ if $i_0\notin \hu\backslash\bu$ or $i_1\notin \hu\backslash\bu$
\item Abort if $(i_0,\eid,*,*)\in\isl$ or $(i_1,\eid,*,*)\in\isl$
\item $\sigma^*\rand \gsig(\gpk,\gsk[i_b],\eid,m)$
\item $\cl := \setd{(i_b,\eid,m,\sigma^*), (i_{1-b},\eid,m,\bot)}$
\item Return $\sigma^*$
\end{itemize}
\end{minipage}
\\
\caption{Oracles used in experiments. These oracles  manipulate the following global variables:
a list {\hu} of honest users, % a list {\cu} of corrupted users  whose personal secret keys have been chosen by the adversary,
a list {\bu} of bad users whose group signing keys and personal secret keys have been revealed to the adversary; a list {\isl} such that $\isl[i]$ contains a group signature obtained on the $i$-th query to the signing oracle,
and a list {\cl} of challenge signatures  obtained from the challenge oracle; a list $\upk$ such that $\upk[i]$ contains the public key of a user $i$; a list $\usk$ such that $\usk[i]$ contains the secret key of a user $i$;
a list $\reg$ such that $\reg[i]$ contains the registration information of group member $i$. The lists $\hu,\bu,\cl$ are initially empty, and all entries of the lists $\isl,\upk,\usk,\reg$ are initialised to be $\bot$. The number of successful $\gsig$ queries is recorded by a global variable $\sn$ and its initial value is $\sn = 1$.}\label{fig:oracles}
\end{figure*}

\begin{figure*}[pt]
\scriptsize
%\hspace*{-10mm}
\begin{minipage}[t]{.48\textwidth}
%\begin{multicols}{2}
${\gcor(1^\lambda):}$
\begin{itemize}[leftmargin=0.16in,topsep=0pt]
\item $(\gpk, \mik, \mok) \rand \setup(1^\lambda)$
\item ${\hu} \rand \emptyset; {\bu} \rand \emptyset; {\isl} \rand \emptyset; {\cl} \rand \emptyset$
\item $(i_0, i_1, \eid, m_0, m_1) \rand \A^{\addu, \rreg}(1^\lambda, \gpk)$
\item If $i_0\notin \hu  \mbox{ return } 0$
\item $\sigma_0\rand\gsig(\gpk, \gsk[i_0], \eid, m_0)$
\item $(j,\pi)\rand \open(\gpk, \reg, \mok, \eid,  m_0, \sigma_0)$
\item If $\gver(\eid, m_0, \sigma_0) = 0$ or $j \neq i_0$ then return 1
\item If $\judge(\gpk,i_0,\upk[i_0],\sigma_0,\pi) =0$ then return 1
\item Parse $\sigma_0$ to extract the event public key $\epk_0$
\item $(\sigma_e,s_e)\rand \esig(\usk[i_0], \eid, \epk_0, m)$
\item If $\ever(\gpk,\eid, \epk_0, m_e,\sigma_e) =0$ then return 1
\item If $i_1\notin \hu$ return 0
\item $\sigma_1\rand\gsig(\gpk, \gsk[i_1], \eid, m_1)$
\item If $i_0 = i_1$ and $\link(\eid, m_0, \sigma_0, m_1, \sigma_1) = 0$ then return 1
\item If $i_0 \neq i_1$ and $\link(\eid, m_0, \sigma_0, m_1, \sigma_1) = 1$ then return 1
\item Return 0
\end{itemize}

\noindent\makebox[\linewidth]{\rule{\textwidth}{0.3pt}}

${\gtr(1^\lambda):}$
\begin{itemize}[leftmargin=0.16in,topsep=0pt]
\item $(\gpk, \mik, \mok) \rand \setup(1^\lambda)$
\item ${\hu} \rand \emptyset; {\bu} \rand \emptyset;   {\isl} \rand \emptyset; {\cl\rand \emptyset}$
\item $(\eid, m, \sigma) \rand \A^{\addu,  \rreg, \ogsk}(1^\lambda, \gpk, \mok)$
\item If ${\gver}(\gpk, \eid, m,\sigma) = 0 \mbox{ then return } 0$
\item $(i,\pi) \rand \open(\gpk, \mok, \reg, \eid, m,\sigma)$
\item If $i = \bot$ or $\judge(\gpk,i,\upk[i],\sigma,\pi) = 0$ then return 1 else return 0
\end{itemize}

\noindent\makebox[\linewidth]{\rule{\textwidth}{0.3pt}}

${\glk(1^\lambda):}$
\begin{itemize}[leftmargin=0.16in,topsep=0pt]
\item $(\gpk, \mik, \mok) \rand \setup(1^\lambda)$
\item ${\hu} \rand \emptyset; {\bu} \rand \emptyset;  {\cu} \rand \emptyset;  {\isl} \rand \emptyset; {\cl\rand \emptyset}$
\item $(\eid, m_0, \sigma_0, m_1, \sigma_1) \rand \A^{\addu,  \rreg, \ogsk}(1^\lambda, \gpk, \mok)$
\item If ${\gver}(\gpk, \eid, m_0,\sigma_0) = 0$ or ${\gver}(\gpk, \eid, m_1,\sigma_1)\\ = 0$ then return 0
\item $(i_b,\pi_b) \rand \open(\gpk, \mok, \reg, \eid, m_b,\sigma_b)$ for $b = 0,1$
\item If $i_0 = \bot$ or $\judge(\gpk,i_0,\upk[i_0],\sigma_0,\pi_0) = 0$ then return 0
\item If $i_1 = \bot$ or $\judge(\gpk,i_1,\upk[i_1],\sigma_1,\pi_1) = 0$  then return 0
\item If $i_0 = i_1 \wedge 0= \link(\eid, m_0, \sigma_0, m_1, \sigma_1)$ then return 1
\item If $i_0\neq i_1 \wedge 1= \link(\eid, m_0, \sigma_0, m_1, \sigma_1)$ then return 1
\item Return 0
\end{itemize}

\end{minipage}%
~~~~\vline~~~~
\begin{minipage}[t]{.48\textwidth}

${\gany(1^\lambda):}$
\begin{itemize}[leftmargin=0.16in,topsep=0pt]
\item $(\gpk, \mik, \mok) \rand \setup(1^\lambda)$
\item ${\hu} \rand \emptyset; {\bu} \rand \emptyset; {\isl} \rand \emptyset; {\cl} \rand \emptyset$
\item $(i_0, i_1, \eid, m, st)\\~~~~~~~~~~~~~~ \rand \A_0^{\sndu,\ogsk, \gsig,  \wreg, \esig}(1^\lambda, \gpk,\mik)$
\item $b \rand \setd{0,1}$; $\sigma^*\rand \ch(i_0, i_1, \eid, m)$
\item $\hat b \rand \A_1^{\sndu,\ogsk, \gsig,  \wreg, \esig}(st, \sigma^*)$
\item Return $\hat b = b$
\end{itemize}

\noindent\makebox[\linewidth]{\rule{\textwidth}{0.3pt}}

${\gnfg(1^\lambda):}$
\begin{itemize}[leftmargin=0.16in,topsep=0pt]
\item $(\gpk, \mik, \mok) \rand \setup(1^\lambda)$
\item ${\hu} \rand \emptyset; {\bu} \rand \emptyset;  {\isl} \rand \emptyset; {\cl} \rand \emptyset$
\item $(\eid, m, \sigma, i,\pi)\\~~~~~~~~~~~~~ \rand \A^{\sndu, \ogsk, \wreg,\gsig,\esig}(1^\lambda, \gpk, \mik, \mok)$
\item Return 1 if the following are all true
\begin{itemize}
\item ${\gver}(\gpk, \eid, m,\sigma) = 1$
\item $i\in \hu \backslash\bu$ and $\judge(\gpk,i,\upk[i],\sigma,\pi) = 1$
\item $\A$ did not query $\gsig(i, \eid, m)$
\end{itemize}
\item Return 0.
\end{itemize}

\noindent\makebox[\linewidth]{\rule{\textwidth}{0.3pt}}

${\gel(1^\lambda):}$
\begin{itemize}[leftmargin=0.16in,topsep=0pt]
\item $(\gpk, \mik, \mok) \rand \setup(1^\lambda)$
\item ${\hu} \rand \emptyset; {\bu} \rand \emptyset;  {\isl} \rand \emptyset; {\cl} \rand \emptyset$
\item $(\eid, \setd{m_b, \sigma_b, i_b,\pi_b}_{b=0,1}) \\ ~~~~~~~~~~~~~\rand \A^{\ogsk, \sndu, \wreg, \gsig,\esig}(1^\lambda, \gpk,\mik,\mok)$
\item If $\gver(\gpk, \eid, m_0, \sigma_0) = 0$ or $\gver(\gpk, \eid, m_1, \sigma_1)\\ = 0$  then return 0
\item If $\judge(\gpk,i_0,\upk[i_0],\sigma_0,\pi_0) = 0$ or $\judge(\gpk,i_1,\\\upk[i_1],\sigma_1,\pi_1) = 0$ then return 0
\item If $i_0\neq i_1$ and $1= \link(\eid, m_0, \sigma_0, m_1, \sigma_1)$ then return 1
\item If $i_0 = i_1$ and $0= \link(\eid, m_0, \sigma_0, m_1, \sigma_1)$ then return 1
\item Return 0
\end{itemize}

\noindent\makebox[\linewidth]{\rule{\textwidth}{0.3pt}}

${\gnfe(1^\lambda):}$
\begin{itemize}[leftmargin=0.16in,topsep=0pt]
\item $(\gpk, \mik, \mok) \rand \setup(1^\lambda)$
\item ${\hu} \rand \emptyset; {\bu} \rand \emptyset;  {\isl} \rand \emptyset; {\cl} \rand \emptyset$
\item $(k,\sigma_e,m_e) \\~~~~~~~~~\rand \A^{\sndu, \ogsk,\wreg,\gsig,\esig}(1^\lambda, \gpk, \mik, \mok)$
\item If $\isl[k] = \bot$  then return  0
\item Parse $\isl[k] = (i,\eid,m,\sigma)$ and extract the event public key $\epk$ from $\sigma$
\item Return 1 if the following are all true 
\begin{itemize}
\item $\ever(\eid, \epk, m_e,\sigma_e) = 1$
\item $\A$  did not query $\esig(k,  m_e)$ or $\ogsk(i)$
\end{itemize}
\item Return 0
\end{itemize}

%\noindent\makebox[\linewidth]{\rule{\textwidth}{0.3pt}}

\end{minipage}
\\
\\
\caption{Experiments for correctness, anonymity, traceability, event linkability and unforgeability. Formally, 
an {\name}  protocol is {\em correct} if the advantage $\advc(1^\lambda) := \pr[\gcor(1^\lambda) = 1]$ is negligible. %in $\lambda$ for any PPT adversary $\A$.  
An {\name}  protocol is {\em anonymous} if the advantage $\adva(1^\lambda):= \abt{\pr[\gany(1^\lambda) = 1] - {1}/{2}}$ is negligible.
% in $\lambda$ for any PPT adversary $\A = (\A_0, \A_1)$. 
An {\name}  protocol is {\em traceable} if  the advantage $\advtr(1^\lambda):= \pr[\gtr(1^\lambda) = 1]$ is negligible.
% in $\lambda$ for any PPT adversary $\A$. 
An {\name}  protocol is {\em event linkable} if the advantage $\advlk(1^\lambda):= \pr[\glk(1^\lambda) = 1]$ is negligible.
% in $\lambda$ for any PPT adversary $\A$. 
An {\name}  protocol is {\em gsig-non-frameable} if  the advantage $\advnfg(1^\lambda):= \pr[\gnfg(1^\lambda) = 1]$ is negligible.
% in $\lambda$ for any PPT adversary $\A$. 
An {\name}  protocol is {\em esig-unforgeable} if the advantage $\advnfe(1^\lambda):= \pr[\gnfe(1^\lambda) = 1]$ is negligible.
% in $\lambda$ for any PPT adversary $\A$.
An {\name}  protocol is {\em link-unforgeable} if for any $\lambda\in\mathbb{N}$, the advantage $\advel(1^\lambda):= \pr[\gel(1^\lambda) = 1]$ is negligible. % in $\lambda$ for any PPT adversary $\A$.
}\label{fig:games}
\end{figure*}

\begin{proof}{Theorem \ref{thm:anony}}
Assume an adversary $\A$ breaks  the anonymity of the {\name} protocol. We shall construct an adversary $\B$ which breaks the XDH assumption. The proof proceeds with a sequence of games.
\begin{description}[leftmargin=*]
  \item[{\bf Game 0.}] Let Game 0 be the original game $\gany(1^\lambda)$ and let $S_0$ be the event that $\hat b = b$ in this game. Clearly we have $\adva(1^\lambda) = \abt{\pr[S_0]-1/2}$.
      \item[{\bf Game 1.}] Let Game 1 be the  same as Game 0, except $B\rand \mathbb{G}_1$ in the challenge signature $\sigma^*$ is chosen uniformly at random from $\mathbb{G}_1$. Let $S_1$ be the event that $\hat b = b$ in this game. We shall prove that $\abt{\pr[S_1]-\pr[S_0]} \leq \negl(\lambda)$  under the XDH assumption in the random oracle model.

      Suppose there is a poly-time attacker $\A$ that has non-negligible difference between $S_0$ and $S_1$. Then we shall construct a poly-time algorithm $\B$ that breaks the XDH assumption.

      Let $({\mathbb G}_1, p, g, g^{\xi}, g^{\alpha}, g^{\beta})$ be a XDH problem given to $\B$. $\B$'s goal is to decide whether $\beta = \xi \alpha$. ${\mathbb G}_1$ is assumed to be associated with a bilinear map $\pair: {\mathbb G}_1\times {\mathbb G}_2 \rightarrow {\mathbb G}_T$.

$\B$ generates parameters for the {\name} scheme. Choose $g_1 \rand \mathbb{G}_1$, and $\gamma\rand \mathbb{Z}^*_p$, and $g_2\rand \mathbb{G}_2$, and $\hs: \setd{0,1}^*\rightarrow \mathbb{G}_1$ and $\hsh: \setd{0,1}^*\rightarrow \mathbb{Z}_p^*$. Set $u = g, h = g^{\xi} (=  u^{\xi}), w = g_2^{\gamma}$.  The group public key is $\gpk = (g_1, h, u,\hs, \hsh, g_2, w)$, the master issuing key is $\mik = \gamma$ and the master opening key is $\mok = \xi$. $\B$ gives $\gpk$ and $\mik$ to $\A$. Note that the master opening key is $\xi$ which is unknown to $\B$.
Since $\B$ generates the master issuing key $\mik$, it can answer $\A$'s queries on $\sndu, \ogsk,  \wreg,\esig$ easily and the details are omitted here. Other queries are answered as below.
\begin{itemize}[leftmargin=*]
\item On a hash query $\eid\in\setd{0,1}^*$ to $\hs$,  if $\eid$ has not been queried before, then choose a random  $r\rand\mathbb{G}_1$, set $\hs(\eid) = r$ and return $r$ to $\A$.
\item On a hash query $M\in\setd{0,1}^*$ to $\hsh$,  if $M$ has not been queried before, then choose a random  $r\rand\mathbb{Z}_p^*$, set $\hsh(M) = r$ and return $r$ to $\A$.
\item On the challenge query, $\A$ submits two indices $i_0, i_1$, an event $\eid$, a message $m$. $\B$ chooses a random bit $b\rand\setd{0,1}$. Let $\gsk[i_b] =(x,y,A)$ be the user $i_b$'s group signing key. Then $\B$ computes $D = g^{\alpha}, B = A g^{\beta}, T = \hs(\eid)^y$, choose randomly $c, s_x , s_y,  s_\alpha, s_\delta \rand \mathbb{Z}^*_p$ and then compute ${R}_1 = u^{s_\alpha} D^{c}, R_2 = {\hs(\eid)}^{s_y} T^{-c}, R_3 = u^{s_\delta} D^{s_x}$ and 
$R_4 = \pair(B, g_2)^{s_x}\pair(h, w)^{s_\alpha}\pair(h,g_2)^{s_y+s_\delta}\bt{\frac{\pair(B, w)}{\pair(g_1, g_2)}}^c$. $\B$ sets $\hsh(\eid, m, D, B, T, R_1, R_2, R_3,R_4) = c$. %It encounters a collision only with negligible probability. In case of a collision, $\B$ aborts. Otherwise
$\B$ returns a signature $\sigma^* = (D,B,T,c,s_x, s_y,s_\alpha,s_\delta)$ as a challenge.
\end{itemize}
$\A$ outputs a guess bit $\hat b$ and $\B$ outputs $\hat b = b$ as a solution to the XDH problem.
In this simulation, when $\beta = \xi \alpha$, $\sigma^*$ is a valid signature. $\A$ plays against Game 0 % when $\B$ does not abort the game,
and therefore $\pr[\B({\mathbb G}_1, p, g, g^{\xi}, g^{\alpha}, g^{\xi \alpha}) = 1] = \pr[S_0]$. When $\beta$ is a uniform random, $B$ in $\sigma^*$ is uniformly distributed. Hence $\A$ plays against Game 1 and $\pr[\B({\mathbb G}_1, p, g, g^{\xi}, g^{\alpha}, g^{\beta}) = 1] = \pr[S_1]$. By XDH assumption, we know
$\abt{\pr[S_1] - \pr[S_0]} \leq |
\pr[\B({\mathbb G}_1, p, g, g^{\xi}, g^{\alpha}, g^{\xi\alpha}) = 1] - \pr[\B({\mathbb G}_1, p, g, g^{\xi}, g^{\alpha}, g^{\beta}) = 1]| \leq \negl(\lambda)$.
          \item[{\bf Game 2.}] Let Game 2 be the same as Game 1, except $T\rand \mathbb{G}_1$  in the challenge signature $\sigma^*$ is chosen uniformly at random from $\mathbb{G}_1$. Let $S_2$ be the event that $\hat b = b$ in this game.
We shall prove $\abt{\pr[S_2] - \pr[S_1]} \leq \negl(\lambda)$  under the XDH assumption in the random oracle model.
      Suppose there is a poly-time attacker $\A$ that has a non-negligible difference between $S_1$ and $S_2$. Then we shall construct a poly-tme algorithm $\B$ that breaks the XDH assumption. 
      Let $({\mathbb G}_1, p, g_1, g_1^{y}, g_1^{\eta}, g_1^{\beta})$ be a XDH problem given to $\B$. $\B$'s goal is to decide whether $\beta = y\eta$. ${\mathbb G}_1$ is assumed to be associated with a bilinear map $\pair: {\mathbb G}_1\times {\mathbb G}_2 \rightarrow {\mathbb G}_T$.

$\B$ generates parameters for the {\name} scheme. $\B$ chooses $\xi, \gamma, a \rand \mathbb{Z}^*_p, g_2\rand \mathbb{G}_2$, and $\hs: \setd{0,1}^*\rightarrow \mathbb{G}_1$ and $\hsh: \setd{0,1}^*\rightarrow \mathbb{Z}_p$ and sets $u = g_1^a, h = u^{\xi}, w = g_2^{\gamma}$.  The  group public key $\gpk = (g_1, h, u, \hs, \hsh, g_2, w)$. $\B$ gives $\gpk$ and $\mik$ to $\A$. Suppose $\A$ queries to add at most $q_A$ users and queries at most $q_\hs$ (distinct) hashes to $\hs$ (this also includes the $\hs$ query in $\gsig$ and $\ch$), $\B$ chooses $i_0^*, i_1^*\rand \setd{1,\cdots, q_A}$ and $j^*\rand \setd{1,\cdots, q_\hs}$ uniformly at random. $\B$ fixes $i_0^*, j^*$ as the challenge query. Note that it does not matter whether $i_0^* = i_1^*$.
\begin{itemize}[leftmargin=*]
\item On the queries to the hash function $\hs$, $\B$ maintains a list $L$. On the $j$-th distinct hash query $\eid\in\setd{0,1}^*$ to $\hs$, if $\eid$ has not queried before, then
\begin{itemize}[leftmargin=0.08in]
\item if $j = j^*$, set $L = L \cup(\eid, j, g_1^{\eta})$ and $\hs(\eid) = g_1^{\eta}$. $\B$ returns $g_1^\eta$ to $\A$.
\item Otherwise choose a random $r\rand\mathbb{Z}_p^*$, set $L = L \cup(\eid, j, r, g_1^r)$ and $\hs(\eid) = g_1^r$. $\B$ returns $g_1^r$ to $\A$.
\end{itemize}

\item On a hash query $M\in\setd{0,1}^*$ to $\hsh$, if $M$ has not been queried before, then choose a random $r\rand\mathbb{Z}_p^*$ and set $\hsh(M) = r$.
\item On a $\sndu(i,M_{in})$ query, if $i\in \hu$, then $\B$ return $\bot$. Otherwise update $\hu = \hu\cup\setd{i}$ and compute group signing key for user $i$ as follows:
\begin{itemize}[leftmargin=0.08in]
\item If $i\neq i_0^*$, choose $x_i, y_i\rand \mathbb{Z}^*_p$, compute $A_i = (g_1 h^{-y_i})^{\frac{1}{\gamma+x_i}}$ and set $\gsk[i] = (x_i,y_i,A_i)$.
\item If $i = i_0^*$, choose $x\rand \mathbb{Z}^*_p$ and compute $A = (g_1 (g_1^y)^{-a\xi})^{\frac{1}{\gamma+x}} = (g_1 h^{-y} )^{\frac{1}{\gamma+x}}$ and set $\gsk[i^*] = (x,\bot,A)$ because $\B$ does not know $y$.
\end{itemize}
\item On a $\ogsk(i)$ query, if $i \in \setd{i_0^*,i_1^*}$, then $\B$ aborts. This corresponds to the definition of  $\ch$, which requires the group signing keys of the users queried in the challenge phase have not been revealed by any  $\ogsk(i)$ query. Otherwise if $i\neq i^*$, $\B$ returns $\gsk[i]$ to $\A$.
\item On a $\gsig(i,\eid,m)$ query, if $i\notin \hu$ or $\gsk[i] = \bot$ then $\B$ returns $\bot$. If $\eid$ has not been queried to $\hs$ before, $\B$ first simulates the query to $\hs$ as described above. Assume $\eid$ corresponds to $j$-th distinct query to $\hs$. Then $\B$ computes a valid signature as follows:
\begin{itemize}[leftmargin=0.08in]
\item If $i\notin \setd{i_0^*,i_1^*}$ and $j\neq j^*$, assume $\gsk[i] = (x_i,y_i,A_i)$ and $(\eid, j, r,g_1^r)\in L$. Choose $B\rand \mathbb{G}_1$ and $\alpha, c, s_{x_i}, s_{y_i}, s_\alpha, s_\delta \rand\mathbb{Z}^*_p$.  Compute $D = u^\alpha, T = g_1^{r y_i}, {R}_1 = u^{s_\alpha} D^{c}, R_2 = {\hs(\eid)}^{s_y} T^{-c}, R_3 = u^{s_\delta} D^{s_x}$ and $R_4 = \pair(B, g_2)^{s_x}\pair(h, w)^{s_\alpha}\pair(h,g_2)^{s_y+s_\delta}\bt{\frac{\pair(B, w)}{\pair(g_1, g_2)}}^c$. Then set $\hsh(\eid, m, D, B, T, R_1, R_2, R_3, R_4) = c$. $\B$ returns a signature $\sigma = (D,B,T,c, s_{x_i}, s_{y_i}, s_\alpha, s_\delta)$.
\item If $i\notin \setd{i_0^*,i_1^*}$ and $j = j^*$, assume $\gsk[i] = (x_i,y_i,A_i)$ and $(\eid, j, g_1^\eta)\in L$. Choose $\alpha, c, s_{x_i}, s_{y_i}, s_\alpha, s_\delta \rand\mathbb{Z}^*_p$ and $B\rand \mathbb{G}_1$.  Compute $D = u^\alpha, T = g_1^{\eta y_i}, {R}_1 = u^{s_\alpha} D^{c}, R_2 = {\hs(\eid)}^{s_y} T^{-c}, R_3 = u^{s_\delta} D^{s_x}, R_4 = \pair(B, g_2)^{s_x}\pair(h, w)^{s_\alpha}\pair(h,g_2)^{s_y+s_\delta}\bt{\frac{\pair(B, w)}{\pair(g_1, g_2)}}^c$.  Then set $\hsh(\eid, m, D, B, T, R_1, R_2, R_3, R_4) = c$. $\B$ returns a signature $\sigma = (D,B,T,c, s_{x_i}, s_{y_i}, s_\alpha, s_\delta)$.
\item If $i = i_0^*$ and $j\neq j^*$, assume $\gsk[i] = (x,\bot,A)$ and $(\eid, j, r,g_1^r)\in L$. In this case, $\B$ does not have $y$, but $\B$ can simulate a signature by applying $r$ to $g_1^y$. Choose $\alpha, c, s_{x}, s_{y}, s_\alpha, s_\delta \rand\mathbb{Z}^*_p$ and $B\rand \mathbb{G}_1$.  Compute $D = u^\alpha, T = g_1^{r y}, {R}_1 = u^{s_\alpha} D^{c}, R_2 = {\hs(\eid)}^{s_y} T^{-c}, R_3 = u^{s_\delta} D^{s_x}, R_4 = \pair(B, g_2)^{s_x}\pair(h, w)^{s_\alpha}\pair(h,g_2)^{s_y+s_\delta}\bt{\frac{\pair(B, w)}{\pair(g_1, g_2)}}^c$.  Then set $\hsh(\eid, m, D, B, T, R_1, R_2, R_3, R_4) = c$. $\B$ returns a signature $\sigma = (D,B,T,c, s_{x}, s_{y}, s_\alpha, s_\delta)$.
\item If $i = i_1^*$ and $j\neq j^*$, assume $\gsk[i] = (x_i,y_i,A_i)$ and $(\eid, j, r,g_1^r)\in L$.  Choose $\alpha, c, s_{x_i}, s_{y_i}, s_\alpha, s_\delta \rand\mathbb{Z}^*_p$ and $B\rand \mathbb{G}_1$.  Compute $D = u^\alpha, T = g_1^{r y_i}, {R}_1 = u^{s_\alpha} D^{c}, R_2 = {\hs(\eid)}^{s_y} T^{-c}, R_3 = u^{s_\delta} D^{s_x}, R_4 = \pair(B, g_2)^{s_x}\pair(h, w)^{s_\alpha}\pair(h,g_2)^{s_y+s_\delta}\bt{\frac{\pair(B, w)}{\pair(g_1, g_2)}}^c$.  Then set $\hsh(\eid, m, D, B, T, R_1, R_2, R_3, R_4) = c$. $\B$ returns a signature $\sigma = (D,B,T,c, s_{x_i}, s_{y_i}, s_\alpha, s_\delta)$.
\item If $i\in \setd{i_0^*,i_1^*}$ and $j=j^*$, then $\B$ aborts because this query is supposed to be the challenge query. Therefore $\B$ aborts. Note that this corresponds to the definition of $\ch$, which requires the challenge queries  to have not been queried to  $\gsig$ before.
\end{itemize}
\item On a query $\wreg(i,val)$, $\B$ sets $\reg[i] = val$.
\item On the challenge query, $\A$ submits two indices $i_0, i_1$, an event $\eid$, a message $m$.
If $\eid$ has not been queried to $\hs$ before, $\B$ first simulates the query to $\hs$ as described above. Then $\B$ chooses a random bit $b\rand\setd{0,1}$. If $i_b \neq i_0^*$ or $i_{1-b}\neq i_1^*$ or $\eid$ is not the $j^*$-the distinct query to $\hs$, then $\B$ aborts and outputs {\sf Fail}. Since $\gsk[i_0^*] =(x,\bot,A)$ and $(\eid, j, g_1^\eta )\in L$, $\B$ computes the challenge signature as follows. $\B$ chooses $\alpha, c, s_x , s_y,  s_\alpha, s_\delta \rand \mathbb{Z}^*_p$ and $B \rand \mathbb{G}_1$, computes $D = u^{\alpha}, T = g_1^\beta,{R}_1 = u^{s_\alpha} D^{c}, R_2 = {\hs(\eid)}^{s_y} T^{-c}, R_3 = u^{s_\delta} D^{s_x}, R_4 = \pair(B, g_2)^{s_x}\pair(h, w)^{s_\alpha}\pair(h,g_2)^{s_y+s_\delta}\bt{{\pair(B, w)}/{\pair(g_1, g_2)}}^c$ and sets $\hsh(\eid, m, D, B, T, R_1, R_2, R_3,R_4)=c$.  $\B$ returns a signature $\sigma^* = (D,B,T,c,s_x, s_y,s_\alpha,s_\delta)$ to $\A$ as the challenge. $\B$ sets $\cl = \{(i_b,\eid,m,\sigma^*),\\ (i_{1-b},\eid,m,\bot)\}$. $\A$ outputs $\hat b$. Finally, $\B$ outputs $b=\hat b$.
\item On a $\esig(k,m_e)$ query,
\begin{itemize}[leftmargin=0.08in]
\item If $\isl[k] = (i,\eid,m,\sigma)$ for $k\ge 1$, then parse $\sigma = (D,B,T,c,s_x,s_y,s_\alpha,s_\delta)$. Choose $s_e, c_e\rand \mathbb{Z}^*_p$, compute $R = \hs(\eid)^{s_e} T^{-c_e}$ and set $\hsh(\eid,m,T,R) = c_e$. $\B$ returns $\sigma_e = (s_e,c_e)$.
\item If $k=0$, the adversary queries to get event signatures following the challenge signature $\sigma^*$. Assume $(i_b,\eid,m,\sigma^*)\in \cl$ with $\sigma^*\neq\bot$. Since $\hs(\eid) = g_1^\eta$, $\B$ chooses $s_e,c_e\rand \mathbb{Z}^*_p$ and computes $R = g_1^{s_e\eta}g_1^{-c_e \beta}$ and set $\hsh(\eid, m, g_1^{\eta}, R) = c_e$. $\B$ returns $\sigma_e  = (s_e,c_e)$ to $\A$.
\end{itemize}
\end{itemize}
%Similarly, if any collision to the hash queries occurs, then $\B$ aborts and this only happens with negligible probability.
Since $i_0^*, i_1^*,j^*$ are uniformly  and independently drawn, the probability that {\sf Fail} does not occur is $\frac{1}{q_A^2 q_{\hs}}$.
 \begin{itemize}[leftmargin=*]
 \item If $\B$ is given $({\mathbb G}_1, p, g_1, g_1^{y}, g_1^{\eta}, g_1^{\beta})$ with $\beta = y \eta$, then $\B$ simulates Game 1 perfectly when {\sf Fail} does not occur. It is easy to see  $ \pr[\B({\mathbb G}_1, p, g_1, g_1^{y}, g_1^{\eta}, g_1^{y \eta}) = 1] = \frac{1}{q_A^2 q_{\hs}}\pr[S_1] $.
 \item If $\B$ is given $({\mathbb G}_1, p, g_1, g_1^{y}, g_1^{\eta}, g_1^{\beta})$ with $\beta \rand \mathbb{Z}^*_p$, then $\B$ simulates Game 2 perfectly when {\sf Fail} does not occur. It is easy to see
$\pr[\B({\mathbb G}_1, p, g_1, g_1^{y}, g_1^{\eta}, g_1^{\beta}) = 1] = \frac{1}{q_A^2 q_{\hs}}\pr[S_2]$. In this case, all the elements in $\sigma^*$ are independently and randomly drawn, and $\B$'s answers to $\A$'s queries contain no info about $b$. Therefore $\pr[S_2] = 1/2$. Then
$ \abt{\pr[S_2]-\pr[S_1]} \leq {q_A^2 q_{\hs}}  |\pr[\B({\mathbb G}_1, p, g_1, g_1^{y}, g_1^{\eta}, g_1^{y \eta}) = 1] - \pr[\B({\mathbb G}_1, p, g_1, g_1^{y}, g_1^{\eta}, g_1^{\beta}) = 1]|
 \leq \negl(\lambda)$.
\end{itemize}

Finally, from $\adva(1^\lambda) = \abt{\pr[S_0]-1/2}$ and
$\abt{\pr[S_1]-\pr[S_0]} \leq \negl(\lambda)$  and  $\abt{\pr[S_2]-\pr[S_1]}\leq \negl(\lambda)$ and $\pr[S_2] = 1/2$, we know that
$\adva(1^\lambda) \leq \abt{\pr[S_0]-\pr[S_1]} + \abt{\pr[S_2]-\pr[S_1]}\leq \negl(\lambda)$.
\end{description}

\end{proof}

%\begin{theorem}\label{theorem:trace}
%The $\elgs$ scheme is traceable under the $q$-SDH assumption.
%\end{theorem}

%\begin{thmagain}{theorem:trace}
%The $\elgs$ scheme is traceable under the $q$-SDH assumption.
%\end{thmagain}

\begin{proof}{Theorem \ref{theorem:trace}}
Assume an adversary $\A$ breaks the traceability of the {\name} scheme.
We shall construct an adversary $\B$ which breaks the $q$-SDH assumption, where $q$ is the upper bound of the number of group members added by $\A$ through the Oracle queries.

Let $(g'_1,{g_1'}^\gamma,\cdots,{g_1'}^{\gamma^q}, g'_2, {g'_2}^\gamma)$  be a random instance of the $q$-SDH problem. The goal of $\B$ is to find a pair $({g_1'}^{\frac{1}{\gamma+x}},x)$ for some $x\in\mathbb{Z}^*_p$.
$\B$ generates the parameters of {\name} as follows. Choose  $x_i,y_i\rand\mathbb{Z}^*_p$ for $i=1,\cdots,q$. Define the following polynomials,
{\small
\[
\begin{split}
F(X) & = \prod_{j=1}^q (X+x_j) = \sum^q_{k=0} a_k X^k\\
F_i(X)& = F(X)/(X+x_i) = \prod_{j=1, j\neq i}^q (X+x_j) = \sum^{q-1}_{k=0} a_{i,k} X^k \\
F_{i,j}(X)& = \frac{F(X)}{(X+x_i)(X+x_j)} \\
&= \prod_{k=1, k\neq i,j}^q (X+x_j) = \sum^{q-2}_{k=0} a_{i,j,k} X^k,  \mbox{ for } i\neq j
\end{split}
\]}
where coefficients can be efficiently computed using $x_1,\cdots,x_q$. Choose $\eta,\theta_1,\cdots, \theta_q, \beta \rand \mathbb{Z}^*_p$ uniformly.  Then set the parameters as follows: $g_1 = (\prod_{i=0}^q {g_1'}^{a_i \gamma^i})^\eta \prod_{i=1}^q (\prod_{j=0,j\neq i}^{q-1} {g_1'}^{a_{i,j} \gamma^j})^{\theta_i y_i} = {g_1'}^{\eta F(\gamma) +  \sum_{i=1}^q  \theta_i y_i F_i(\gamma)}$, $h = \prod_{i=1}^q (\prod_{j=0,j\neq i}^{q-1} {g'_1}^{a_{i,j} \gamma^j})^{\theta_i} = {g'_1}^{ \sum_{i=1}^q \theta_i F_i(\gamma)}$, $u= h^{1/\xi}$, $g_2 = {g'_2}^\beta$ and $w={g'_2}^{\beta\gamma}$. $\B$ chooses  $\hs: \setd{0,1}^*\rightarrow \mathbb{G}_1$ and $\hsh: \setd{0,1}^*\rightarrow \mathbb{Z}_p$.  The group public key $\gpk = (g_1, h, u, \hs, \hsh, g_2, w)$. The master issuing key $\mik = \gamma$ is unknown to $\B$. The master opening key is $\mok = \xi$. $\B$ initialises the lists $\hu,\bu,\cl$ to be empty, and all entries of the lists $\isl,\upk,\usk,\reg$ to be $\bot$ and $\sn = 1$.  $\B$ answers $\A$'s queries as below:
\begin{itemize}[leftmargin=*]
\item  On a hash query $\eid\in\setd{0,1}^*$ to $\hs$,  if $\eid$ has not been queried before, then choose a random  $r\rand\mathbb{G}_1$, set $\hs(\eid) = r$ and return $r$ to $\A$.
\item On a hash query $M\in\setd{0,1}^*$ to $\hsh$,  if $M$ has not been queried before, then choose a random  $r\rand\mathbb{Z}_p^*$, set $\hsh(\eid) = r$ and return $r$ to $\A$.
\item On an $\addu(i)$ query, $\B$ checks if $i\in \hu$. If not, $\B$ sets $\hu = \hu\cup \setd{i}$ and computes $\upk[i] = h^{y_i}$ and $\usk[i]=y_i$. Since
$g_1 h^{-y_i}   = {g_1'}^{\eta F(\gamma) +  \sum_{k=1}^q \theta_k y_k F_k(\gamma)} {g'_1}^{- y_i  \sum_{k=1}^q \theta_k F_k(\gamma)}   =  {g_1'}^{\eta F(\gamma) + \sum^q_{k=1,k\neq i} \theta_k (y_k-y_i) F_k(\gamma)}$,
$\B$ can compute $A_i = (g_1 h^{-y_i}
)^{\frac{1}{\gamma + x_i}}$ by
\[
\begin{split}
A_i & = \bt{\prod_{k=0}^{q-1}{g'_1}^{a_{i,k}\gamma^k}}^{\eta} \prod^q_{k=1,k\neq i} (\prod_{j=0}^{q-2}{g'_1}^{a_{k,i,j}\gamma^j})^{\theta_k(y_k-y_i)} \\
& = {g_1'}^{\eta  F_i(\gamma) + \sum^q_{k=1,k\neq i} \theta_k (y_k-y_i) F_{k,i}(\gamma)}  = (g_1 h^{-y_i})^{\frac{1}{\gamma + x_i}}
\end{split}
\] $\B$ records $\gsk[i] = (x_i,y_i,A_i)$ and $\reg[i] = (x_i,A_i)$. $\B$ returns $\upk[i]$ to $\A$.
\item The other queries $\rreg,\ogsk$ can be easily answered and the details are omitted here.
\end{itemize}
Finally $\A$ outputs a forge group signature $\sigma = (D,B,T,c,s_x,s_y,s_\alpha,s_\delta)$ on an event $\eid$ and a message $m$. Assume $\gver(\gpk,\eid,m,\sigma) = 1$. Using the forking lemma to rewind $\A$, we can obtain $\sigma' = (D,B,T,c',s'_x,s'_y,s'_\alpha,s'_\delta)$ with $c\neq c'$ with non-negligible probability. Let $\Delta x = \frac{s_x - s'_x}{c-c'}, \Delta y = \frac{s_y - s'_y}{c-c'}, \Delta \alpha = \frac{s_\alpha - s'_\alpha}{c'-c}$.  Dividing two instances of $R_1,R_2,R_3$ we can obtain $D = u^{\Delta \alpha} = u^{\frac{s_\delta - s'_\delta}{s'_x-s_x}}$ and $T = \hsh(\eid)^{\Delta y}$. Therefore we can deduce
${\Delta \alpha} \Delta x = \frac{s_\delta - s'_\delta}{c'-c}$.
Dividing two instances of $R_4$, we can obtain
$
\pair(B, g_2)^{\Delta x}\pair(h, w)^{-\Delta \alpha}\pair(h,g_2)^{\Delta y -\Delta\alpha \Delta x} = \frac{\pair(g_1, g_2)}{\pair(B, w)}
$. Let $A = B h^{-\Delta \alpha}$. By using the above equation, we can easily check $\pair(A,w g_2^{\Delta x}) = e(g_1 h^{-\Delta y}, g_2)$. Therefore $A = (g_1 h^{-\Delta y})^{\frac{1}{\gamma + \Delta x}}$.  Note that the way we compute $A$ indicates that $A$ is the result of the decryption performed in $\open$.
%\begin{itemize}
\begin{description}[leftmargin=*]
\item[Case 1] If $A\notin\setd{A_1,\cdots,A_q}$ and $\Delta x\notin\setd{x_1,\cdots,x_q}$, then the output of $\open$ is $(\bot,\bot)$. We can obtain a new $q$-SDH pair as follows. From $A  = (g_1 h^{-\Delta y})^{\frac{1}{\gamma + \Delta x}}$, we have
$A  = (g_1 h^{-\Delta y})^{\frac{1}{\gamma + \Delta x}}  = {g_1'}^{\frac{\eta F(\gamma) + \sum_{k=1}^q \theta_k (y_k-\Delta y) F_k(\gamma)}{\gamma + \Delta x}}$.
Let $Q(X) = \eta F(X) +  \sum_{k=1}^q \theta_k (y_k-\Delta y) F_k(X)$. We can easily see that $Q(X) = (X + \Delta x) Q'(X) + Q(-\Delta x)$ where $Q(-\Delta x)\in\mathbb{Z}_p$. Since
\[
\small
\begin{split}
& Q(-\Delta x)  = \eta F(-\Delta x) +  \sum_{k=1}^q \theta_k (y_k-\Delta y) F_k(-\Delta x) \\
& = \eta \prod^q_{i=1}(x_i - \Delta x) +  \sum_{k=1}^q \bt{\theta_k (y_k-\Delta y) \prod_{j=1,j\neq k} (x_j-\Delta x)}
\end{split}
\] Because $\Delta x\notin\setd{x_1,\cdots,x_q}$ and $\eta,\theta_1, \cdots, \theta_q$ are chosen independently and randomly, we know $Q(-\Delta x) = 0$ with only negligible probability, no matter what the value $\Delta y$ is. When $Q(-\Delta x) \neq 0$, let $Q'(X) = \sum_{i=0}^{q-1} b_i X^i$ we compute
$S = \bt{A {g'_1}^{-\sum_{i=0}^{q-1} b_i \gamma^i}}^{\frac{1}{Q(-\Delta x)}} = {g'_1}^{\frac{1}{\gamma + \Delta x}}$. $\B$ outputs $(S,\Delta x)$ as a solution to the $q$-SDH instance.

\item[Case 2]  If $A\notin\setd{A_1,\cdots,A_q}$ but $\Delta x = x_i$ for some $i\in\setd{1,\cdots,q}$, then the output of $\open$ is $(\bot,\bot)$.  We can obtain a new $q$-SDH pair as follows.
\[
\begin{split}
A & = (g_1 h^{-\Delta y})^{\frac{1}{\gamma + \Delta x}}
 = {g_1'}^{\frac{\eta F(\gamma) +  \sum_{k=1}^q \theta_k (y_k-\Delta y) F_k(\gamma)}{\gamma +  x_i}} \\
 &  = {g_1'}^{\eta F_i(\gamma) + \bt{\sum_{k=1,k\neq i}^q \theta_k (y_k-\Delta y) F_{k,i}(\gamma)} + \frac{\theta_i (y_i-\Delta y) F_{i}(\gamma)}{(\gamma+x_i)} }\\
\end{split}
\] Clearly $F_i(X) = (X+x_i) F'(X) + F_i(-x_i)$ for some $F'(X)$.
If $\Delta y = y_i$ then we should have $A = A_i$ which contradicts the hypothesis that $A\notin\setd{A_1,\cdots,A_q}$. 
Hence $\Delta y \neq y_i$. Let $a* = \theta_i(y_i-\Delta y) F_i(-x_i)$ and we can see $a*\in \mathbb{Z}_p$ and $a^*\neq 0$.  Assume $\eta F_i(\gamma) + \bt{\sum_{k=1,k\neq i}^q \theta_k (y_k-\Delta y) F_{k,i}(\gamma)} + \theta_i (y_i-\Delta y)F'(X) =  \sum_{i=0}^{q-1} b_i X^i$. Compute
$S = \bt{A {g'_1}^{-\sum_{i=0}^{q-1} b_i \gamma^i}}^{\frac{1}{a^*}} = {g'_1}^{\frac{1}{\gamma + x_i}}
$. $\B$ outputs $(S,x_i)$ as a solution to the $q$-SDH instance.
\item[Case 3]  If $A = A_i$ for some $i\in \setd{1,\cdots,q}$ but $i\notin \hu$. This means $\A$ did not query to add the user $i$ and the number of actual group members is less than $q$. In this case, $\open$ outputs $(\bot,\bot)$. From the attacker's point of view, the distribution of $A_i=(g_1 h^{-y_i})^{x_i}$ is uniformly random. Hence the probability the attacker can guess $A_i$ correctly is only negligible.
\item[Case 4] If $A = A_i$ for some $i\in \setd{1,\cdots,q}$ and $i\in \hu$, but the $\judge$ algorithm outputs 0.  Assume the registration table contains an entry $\reg[i] = (x_i,A_i)$ and the user public key $\upk[i] = h^{-y_i}$. Since the $\judge$ algorithm outputs 0, this means $\pair(A, w g_2^{x_i}) \neq \pair(g_1 h^{-y_i}, g_2)$. Hence $A\neq (g_1 h^{-y_i})^{\frac{1}{\gamma + x_i}}$. This contradicts the hypothesis $A = A_i$.
\end{description}

%\end{itemize}
 \end{proof}
%
% \begin{theorem}\label{thm:eventlink}
%The $\elgs$ scheme is event linkable under the $q$-SDH assumption.
%\end{theorem}

% \begin{thmagain}{thm:eventlink}
%The $\elgs$ scheme is event linkable under the $q$-SDH assumption.
%\end{thmagain}

\begin{proof}{Theorem \ref{thm:eventlink}}
Assume an adversary $\A$ breaks the event linkability of the {\name} protocol.
We shall construct an adversary $\B$ which breaks the $q$-SDH assumption, where $q$ is the upper bound of the number of group members added by $\A$ through the Oracle queries.

Let $(g'_1,{g_1'}^\gamma,\cdots,{g_1'}^{\gamma^q}, g'_2, {g'_2}^\gamma)$  be a random instance of the $q$-SDH problem. The goal of $\B$ is to find a pair $({g_1'}^{\frac{1}{\gamma+x}},x)$ for some $x\in\mathbb{Z}^*_p$.
$\B$ generates the parameters of  {\name}  as follows. Choose $x_i,y_i\rand\mathbb{Z}^*_p$  uniformly at random for $i=1,\cdots,q$. $\B$ then generates the group public key $\gpk = (g_1, h, u, \hs, \hsh, g_2, w)$, the master issuing key $\mik = \gamma$ and the master opening key is $\mok = \xi$ in the exactly the same way as in the proof of Theorem \ref{theorem:trace}. $\B$ answers $\A$'s queries also in the same way as described in the proof of Theorem \ref{theorem:trace}. Finally $\A$ outputs two group signatures $\sigma_b = (D_b,B_b,T_b,c_b,s_{x,b},s_{y,b},s_{\alpha,b},s_{\delta,b})$ on an event $\eid$ and a message $m_b$ with $b=0,1$. Because $\gver(\gpk,\eid,m_b,\sigma_b) = 1$ for $b=0,1$. Similar to the proof of Theorem \ref{theorem:trace}, using the forking lemma to rewind $\A$, we can extract $A'_b = B_b h^{-\Delta\alpha_b} =  (g_1 h^{-\Delta y_b})^{\frac{1}{\gamma + \Delta x_b}}$ and $T_b = \hs(\eid)^{\Delta y_b}$ for $b=0,1$. Because of the output $i_b \neq\bot$ of $\open$ for $b=0,1$, then we know $i_0,i_1$ are the honest users in $\hu$.
From the fact that the $\judge$ also succeeds on both $(i_0,\pi_0)$ and $(i_1,\pi_1)$, we know $A'_b = (g_1 h^{-y_{i_b}})^{\frac{1}{\gamma + x_{i_b}}}$ for $b=0,1$. Therefore  for $b=0,1$ we have:
{\small
\begin{equation}\label{eq:elink}
A'_b = (g_1 h^{-\Delta y_b})^{\frac{1}{\gamma + \Delta x_b}} = (g_1 h^{-y_{i_b}})^{\frac{1}{\gamma + x_{i_b}}}
\end{equation}}
\begin{itemize}[leftmargin=*]
\item For any $b\in\setd{0,1}$, if $\Delta x_b\notin \setd{x_1,\cdots,x_q}$, then $\B$ can compute a  solution $({g_1'}^{\frac{1}{\gamma + \Delta x_b}},\Delta x_b)$ of the $q$-SDH instance, by using a similar argument as in Case 1 in the proof of Theorem \ref{theorem:trace}.
\item For any $b\in\setd{0,1}$, if $\Delta x_b =  x_{k_b}$ for some $k_b\in\setd{1,\cdots,q}$ and $k_b\neq {i_b}$. We first show that $\Delta y_b = y_{k_b}$ with only negligible probability. Otherwise $A_{i_b} = A_{k_b}$ but $i_b\neq k_b$. since $x_{i_b}, y_{i_b}, x_k, y_k$ are chosen independently and randomly, $A_{i_b} = A_{k_b}$ only happens with negligible probability. When $\Delta y_b \neq y_{k_b}$, $\B$ can compute a solution $({g_1'}^{\frac{1}{\gamma + x_{k_b}}}, x_{k_b})$ to the $q$-SDH instance, by using a similar argument as in Case 2  in the proof of Theorem \ref{theorem:trace}.
%Let $h = g^z_1$ for some $z\in\mathbb{Z}_p$. Then we can deduce that $g_1^{\frac{1-z \Delta y_b}{\gamma + \Delta x_b}} = g_1^{\frac{1-z y_{i_b}}{\gamma + x_{i_b}}}$. If $\Delta y_b = y_{k_b}$, then   $x_{i_b}, y_{i_b}, x_k, y_k$ are chosen independently and randomly and $A_k = A_{i_b}$ with only  negligible probability. In this case, $\B$ can compute a solution to the $q$-SDH instance.
\item We are left with the last case when $\Delta x_b =  x_{i_b}$ for both $b=0,1$. Then we can easily see that $\Delta y_b = y_{i_b}$ for $b=0,1$ from Equation \ref{eq:elink}. In this case, if $i_{0} = i_1$, we will have $\Delta y_0 = y_{i_0} = y_{i_1} = \Delta y_1$ and therefore $T_0 = \hs(\eid)^{\Delta y_0} =  \hs(\eid)^{\Delta y_1} = T_1$. If $i_{0} \neq i_1$, we will have $\Delta y_0 = y_{i_0} \neq y_{i_1} = \Delta y_1$ and therefore $T_0 = \hs(\eid)^{\Delta y_0} \neq  \hs(\eid)^{\Delta y_1} = T_1$.
This contradicts the hypothesis that $\A$ breaks the event linkability.
\end{itemize}

\end{proof}

% \begin{theorem}\label{thm:gsignonframe}
%The $\elgs$ scheme is gsig-non-frameable under the DL assumption.
%\end{theorem}

% \begin{thmagain}{thm:gsignonframe}
%The $\elgs$ scheme is gsig-non-frameable under the DL assumption.
%\end{thmagain}
%\vspace{-4pt}

\begin{proof}{Theorem \ref{thm:gsignonframe}}
Assume an adversary $\A$ breaks the non-frameability of the  {\name} protocol.
We shall construct an adversary $\B$ which breaks the DL assumption.

Let $({\mathbb G}_1, p, g, g^{y})$ be a DL problem given to $\B$. $\B$'s goal is to find out $y$. ${\mathbb G}_1$ is assumed to be associated with a bilinear map $\pair: {\mathbb G}_1\times {\mathbb G}_2 \rightarrow {\mathbb G}_T$. $\B$ generates parameters for {\name}. $\B$ chooses $\xi, \gamma, a, z \rand \mathbb{Z}^*_p$ and $g_2\rand \mathbb{G}_2$ and two hash functions $\hs: \setd{0,1}^*\rightarrow \mathbb{G}_1$ and $\hsh: \setd{0,1}^*\rightarrow \mathbb{Z}_p$ and sets $g_1 = g^z, u = g^a, h = g^{a\xi}, , w = g_2^{\gamma}$. $\B$ sets the group public key $\gpk = (g_1, h, u, \hs, \hsh, g_2, w)$, the master issuing key $\mik = \gamma$ and the master opening key $\mok = \xi$. $\B$ gives $(\gpk,\mik,\mok)$ to $\A$.
Let $q$ be the upper bound of the number of group members added by $\A$. $\B$ randomly chooses $i^*\rand\setd{1,\cdots,q}$. $\B$ responses to $\A$'s queries as follows.
\begin{itemize}[leftmargin=*]
\item On a query $\sndu(i,M_{in})$, $\B$ responses as follows. If $i\neq i^*$ and $i\notin \hu$, $\B$ chooses $x_i,y_i\rand\mathbb{Z}_p^*$ and sets $\upk[i] = h^y$ and $\usk[i] = y$. Then $\B$ runs $\join$ and sets $\gsk[i]=(x_i,y_i,A_i=(g_1 h^{-y_i})^{\frac{1}{\gamma+x_i}})$ if the verification is successful. If $i = i^*$, $\B$ chooses $x^*\rand \mathbb{Z}_p^*$ and sets $\upk[i] = h^y = g^{y a \xi}$ and $\usk[i] = \bot$ and $\gsk[i] = (x^*,\bot,A^*=(g_1 h^{-y})^{\frac{1}{\gamma + x^*}})$.
\item On the hash query $\hs(\eid)$  with $\eid\in\setd{0,1}^*$, $\B$ randomly chooses $r\rand\mathbb{Z}_p^*$, sets $\hs(\eid) = g^r$ and maintains a list $L = L\cup\setd{(\eid, r, g^r)}$.
\item On the hash query $\hsh(M)$ with $M\in\setd{0,1}^*$, $\B$ randomly chooses $r\rand\mathbb{Z}_p^*$ and sets $\hsh(M) = r$.
\item On a query $\gsig(i,\eid,m)$,
\begin{itemize}[leftmargin=0.08in]
\item If $i\neq i^*$, $\B$ runs $\sigma\rand \gsig(\gpk,\gsk[i],m)$ and maintains a list $\isl[\sn] = (i,\eid,m,\sigma)$ and then updates $\sn = \sn +1$. $\B$ returns $\sigma$ to $\A$.
\item If $i = i^*$, $\B$ simulates a signature $\sigma$ as follows. $\B$ randomly chooses $\alpha, c, s_x, s_y, s_{\alpha}, s_{\delta}\rand \mathbb{Z}_p^*$. Compute $D = u^\alpha$ and $B = A h^\alpha$. Find the entry $(\eid, r, g^r)$ in the list $L$ and set $T = g^{y r}$. Compute ${R}_1 = u^{s_\alpha} D^{c}, R_2 = {\hs(\eid)}^{s_y} T^{-c}, R_3 = u^{s_\delta} D^{s_x}, R_4 = \pair(B, g_2)^{s_x}\pair(h, w)^{s_\alpha}\pair(h,g_2)^{s_y+s_\delta}\bt{\frac{\pair(B, w)}{\pair(g_1, g_2)}}^c$.  Then set $\hsh(D,B,T, R_1, R_2, R_3, R_4) = c$. $\B$ returns $\sigma = (D,B,T,c,s_x, s_y, s_{\alpha}, s_{\delta})$ to $\A$.
\end{itemize}
\item On a query $\esig(k,m_e)$, $\B$ checks the signature list $\isl$ and find $\isl[k] = (i,\eid,m,\sigma)$. Parse $\sigma =(D,B,T, c, s_x, s_y, s_{\alpha}, s_{\delta})$,
\begin{itemize}[leftmargin=0.08in]
\item If $i\neq i^*$, then $\B$ runs $\sigma_e\rand\esig(\usk[i],\eid, T, m_e)$ and returns $\sigma_e$ to $\A$.
\item If $i = i^*$, then $\B$ simulates the signature as follows.
$\B$ randomly chooses $s_e, c_e\rand\mathbb{Z}_p^*$. $\B$ computes $R = \hs(\eid)^{s_e} T^{-c_e}$ and sets $\hsh(\eid,m_e, T, R) = c_e$.
\end{itemize}
\item On a query $\ogsk(i)$, if $i = i^*$ then $\B$ aborts. Otherwise $\B$ returns $(\gsk[i],\usk[i])$ to $\A$.
\item On a query $\wreg(i,val)$, $\B$ sets $\reg[i] = val$.
\end{itemize}

Finally $\A$ output $(\eid,m,\sigma,i,\pi)$. If $i\neq i^*$, $\B$ aborts and output $\fail$.  Since $i^*$ is chosen uniformly at random, the probability that $\fail$ does not happen is  $1/q$.
If $\gsig(i,\eid,m)$ has been queried before, then $\B$ aborts.

Parse $\sigma = (D,B,T,c, s_x,s_y,s_\alpha,s_\delta)$ and $\pi = (K,s_e,c_e,\\x)$ and let $\upk[i^*] = h^y$. Since $\gver(\gpk,\eid,m,\sigma) = 1$, using the forking lemma, we can extract $(\Delta x, \Delta y, A = (g_1 h^{-\Delta y})^{\frac{1}{\gamma+\Delta x}})$ from the signature $\sigma$. Because the $\judge$ algorithm outputs 1, we have $\pair(A, w g_2^x) = \pair(g_1 h^{-y}, g_2)$. This means $A = (g_1 h^{-y})^{\frac{1}{\gamma+x}}$. Hence we have $(g_1 h^{-\Delta y})^{\frac{1}{\gamma+\Delta x}} = (g_1 h^{-y})^{\frac{1}{\gamma+x}}$. Replacing $g_1 = g^z, h = g^{a\xi}$ in this equation we can obtain
$g^y = g^{\frac{1}{a\xi}\bt{z -\frac{(z-\Delta y a\xi)(\gamma + x)}{\gamma + \Delta x}}} = g^{\frac{z(\Delta x - x) + \Delta y a\xi(\gamma+x)}{(\gamma+\Delta x)a\xi}}
$. Since $\B$ knows $z, a, \xi, \gamma, x, \Delta x, \Delta y$,  $\B$ has found a solution for $y$.
\end{proof}

%\begin{theorem}\label{thm:eventunforge}
%The $\elgs$ scheme is event-unforgeable under the DL assumption.
%\end{theorem}

%\begin{thmagain}{thm:eventunforge}
%The $\elgs$ scheme is event-unforgeable under the DL assumption.
%\end{thmagain}

\begin{proof}{Theorem \ref{thm:eventunforge}}
Assume an adversary $\A$ breaks the  event-unforgeability of  the  {\name} protocol.
We shall construct an adversary $\B$ which breaks the DL assumption.

Let $({\mathbb G}_1, p, g, g^{y})$ be a DL problem given to $\B$. $\B$'s goal is to find out $y$. ${\mathbb G}_1$ is assumed to be associated with a bilinear map $\pair: {\mathbb G}_1\times {\mathbb G}_2 \rightarrow {\mathbb G}_T$. $\B$ generates parameters for {\name}. $\B$ chooses $\xi, \gamma, a \rand \mathbb{Z}^*_p$ and $g_1\rand \mathbb{G}_1$ and $g_2\rand \mathbb{G}_2$ and two hash functions $\hs: \setd{0,1}^*\rightarrow \mathbb{G}_1$ and $\hsh: \setd{0,1}^*\rightarrow \mathbb{Z}_p$ and sets $u = g^a, h = g^{a\xi}, , w = g_2^{\gamma}$. $\B$ sets the group public key $\gpk = (g_1, h, u, \hs, \hsh, g_2, w)$, the master issuing key $\mik = \gamma$ and the master opening key $\mok = \xi$. $\B$ gives $(\gpk,\mik,\mok)$ to $\A$.
Let $q$ be the upper bound of the number of group members added by $\A$. $\B$ randomly chooses $i^*\rand\setd{1,\cdots,q}$. $\B$ responses to $\A$'s queries as below.

\begin{itemize}[leftmargin=*]
\item On a query $\sndu(i,M_{in})$, $\B$ responses as follows. If $i\neq i^*$ and $i\notin \hu$, $\B$ chooses $x_i,y_i\rand\mathbb{Z}_p^*$ and sets $\upk[i] = h^{y_i}$ and $\usk[i] = y_i$. Then $\B$ runs $\join$ and sets $\gsk[i]=(x_i,y_i,A_i=(g_1 h^{-y_i})^{\frac{1}{\gamma+x_i}})$ if the verification is successful. If $i = i^*$, $\B$ chooses $x^*\rand \mathbb{Z}_p^*$ and sets $\upk[i] = h^y = g^{y a \xi}$ and $\usk[i] = \bot$ and $\gsk[i] = (x^*,\bot,A^*=(g_1 h^{-y})^{\frac{1}{\gamma + x^*}})$ since $\B$ does not have $y$.
\item On a hash query $\hs(\eid)$  with $\eid\in\setd{0,1}^*$, $\B$ randomly chooses $r\rand\mathbb{Z}_p^*$, sets $\hs(\eid) = g^r$ and maintains a list $L = L\cup\setd{(\eid, r, g^r)}$.
\item On the hash query $\hsh(M)$ with $M\in\setd{0,1}^*$, $\B$ randomly chooses $r\rand\mathbb{Z}_p^*$ and sets $\hsh(M) = r$.
\item On a query $\gsig(i,\eid,m)$,
\begin{itemize}[leftmargin=0.08in]
\item If $i\neq i^*$, $\B$ runs $\sigma\rand \gsig(\gpk,\gsk[i],m)$ and maintains a list $\isl[\sn] = (i,\eid,m,\sigma)$ and then updates $\sn = \sn +1$. $\B$ returns $\sigma$ to $\A$.
\item If $i = i^*$, $\B$ simulates a signature $\sigma$ as follows. $\B$ randomly chooses $\alpha, c, s_x, s_y, s_{\alpha}, s_{\delta}\rand \mathbb{Z}_p^*$. Compute $D = u^\alpha$ and $B = A h^\alpha$. Find the entry $(\eid, r, g^r)$ in the list $L$ and set $T = g^{y r}$. Compute ${R}_1 = u^{s_\alpha} D^{c}, R_2 = {\hs(\eid)}^{s_y} T^{-c}, R_3 = u^{s_\delta} D^{s_x}, R_4 = \pair(B, g_2)^{s_x}\pair(h, w)^{s_\alpha}\pair(h,g_2)^{s_y+s_\delta}\bt{\frac{\pair(B, w)}{\pair(g_1, g_2)}}^c$.  Then set $\hsh(D,B,T, R_1, R_2, R_3, R_4) = c$. $\B$ returns $\sigma = (D,B,T,c,s_x, s_y, s_{\alpha}, s_{\delta})$ to $\A$.
\end{itemize}
\item On a query $\esig(k,m_e)$, $\B$ checks the signature list $\isl$ and find $\isl[k] = (i,\eid,m,\sigma)$. Parse $\sigma =(D,B,T, c, s_x, s_y, s_{\alpha}, s_{\delta})$
\begin{itemize}[leftmargin=0.08in]
\item If $i\neq i^*$, then $\B$ runs $\sigma_e\rand\esig(\eid,\usk[i],m_e)$ and returns $\sigma_e$ to $\A$.
\item If $i = i^*$, then $\B$ simulates the signature as follows.
$\B$ randomly chooses $s_e, c_e\rand\mathbb{Z}_p^*$. $\B$ computes $R = \hs(\eid)^{s_e} T^{-c_e}$ and sets $\hsh(\eid, m_e, T, R) = c_e$.
\end{itemize}
\item On a query $\ogsk(i)$, if $i = i^*$ then $\B$ aborts. Otherwise $\B$ returns $(\gsk[i],\usk[i])$ to $\A$.
\item On a query $\wreg(i,val)$, $\B$ sets $\reg[i] = val$.
\end{itemize}
$\A$ output $(k,m_e,\sigma_e)$. If $\esig(k,m_e)$ has been queried before, then $\B$ aborts. Let $\isl[k] = (i,\eid,m,\sigma)$. If $i\neq i^*$, $\B$ aborts and output $\fail$.   Since $i^*$ is chosen uniformly at random, the probability that $\fail$ does not happen is  $1/q$. Parse $\sigma = (D,B,T,c, s_x,s_y,s_\alpha,s_\delta)$ and assume $(\eid, r, g^r)$ in the list $L$. Since the signature is simulated by $\B$, we know $T = g^{r y}$.  From $\ever(\gpk,\eid,m,\sigma,m_e,\sigma_e) = 1$, using forking lemma on $\sigma_e$, we can obtain $T = \hs(\eid)^{\frac{s_e - s'_e}{c_e'-c_e}}$. Replacing $T$ with $g^{r y}$ and $\hs(\eid)$ with $g^r$, we have $g^y = g^{\frac{s_e - s'_e}{c_e'-c_e}}$. $\B$ outputs $\frac{s_e - s'_e}{c_e'-c_e}$ as a solution to the DL instance.
\end{proof}

% \begin{theorem}\label{thm:linkunforge}
%The $\elgs$ scheme is link-unforgeable under the DL assumption.
%\end{theorem}

% \begin{thmagain}{thm:linkunforge}
%The $\elgs$ scheme is link-unforgeable under the DL assumption.
%\end{thmagain}

\begin{proof}{Theorem \ref{thm:linkunforge}}
Assume an adversary $\A$ breaks the link-unforgeability of  the  {\name} protocol. We shall construct an adversary $\B$ which breaks the DL assumption. 
Let $({\mathbb G}_1, p, g, g^{z})$ be a DL problem given to $\B$. $\B$'s goal is to find out $z$. ${\mathbb G}_1$ is assumed to be associated with a bilinear map $\pair: {\mathbb G}_1\times {\mathbb G}_2 \rightarrow {\mathbb G}_T$. $\B$ generates parameters for {\name} as follows. $\B$ chooses $\xi, \gamma, a \rand \mathbb{Z}^*_p$ and $g_2\rand \mathbb{G}_2$ and two hash functions $\hs: \setd{0,1}^*\rightarrow \mathbb{G}_1$ and $\hsh: \setd{0,1}^*\rightarrow \mathbb{Z}_p$ and sets $g_1 = g^a,  h = g^{a z}, u = h^{1/\xi}, w = g_2^{\gamma}$. $\B$ sets the group public key $\gpk = (g_1, h, u, \hs, \hsh, g_2, w)$, the master issuing key $\mik = \gamma$ and the master opening key $\mok = \xi$. $\B$ gives $(\gpk,\mik,\mok)$ to $\A$. $\B$ responses queries as follows.
\begin{itemize}[leftmargin=*]
\item On a query $\sndu(i,M_{in})$, if  $i\notin \hu$, $\B$ chooses $x_i,y_i\rand\mathbb{Z}_p^*$ and sets $\upk[i] = h^{y_i}$ and $\usk[i] = y_i$. Then $\B$ runs $\join$ and sets $\gsk[i]=(x_i,y_i,A_i=(g_1 h^{-y_i})^{\frac{1}{\gamma+x_i}})$ if the verification succeeds.
\item On the hash query $\hs(\eid)$ with $\eid\in\setd{0,1}^*$, $\B$ randomly chooses $r\rand\mathbb{Z}_p^*$, sets $\hs(\eid) = r$.
\item On the hash query $\hsh(M)$ with $M\in\setd{0,1}^*$, $\B$ randomly chooses $r\rand\mathbb{Z}_p^*$ and sets $\hsh(M) = r$.
\item On a query $\gsig(i,\eid,m)$, $\B$ checks if $i\in\hu$ and $\gsk[i]\neq \bot$. If successful, $\B$ runs $\sigma\rand \gsig(\gpk,\gsk[i],m)$, sets $\isl[\sn] = (i,\eid,m,\sigma)$ and updates $\sn = \sn +1$. $\B$ returns $\sigma$ to $\A$.
\item On a query $\esig(k,m_e)$, $\B$ checks the signature list $\isl$ and find $\isl[k] = (i,\eid,m,\sigma)$. Parse $\sigma =(D,B,T, c, s_x, s_y, s_{\alpha}, s_{\delta})$. $\B$ runs $\sigma_e\rand\esig(\usk[i],\eid, m_e)$ and returns $\sigma_e$ to $\A$.
\item On a query $\ogsk(i)$, $\B$ returns $(\gsk[i],\usk[i])$ to $\A$.
\item On a query $\wreg(i,\rho)$, $\B$ sets $\reg[i] = \rho$.
\end{itemize}
$\A$ outputs $(\eid, m_0, \sigma_0, i_0,\pi_0, m_1, \sigma_1,i_1,\pi_1)$. Note that,  for $b=0,1$, we have $ \gver(\gpk, \eid, m_b, \sigma_b) = 1$ and $\judge(\gpk,i_b,\upk[i_b],\sigma_b,\pi_b) = 1$ by the definition of ${\gel(1^\lambda)}$. Parse $\sigma_b = (D_b,B_b,T_b, c_b, s_{x,b},s_{y,b},s_{\alpha,b},\\s_{\delta,b})$ and $\pi_b = (K_b,s_b,c_b,x_b')$.
\begin{itemize}[leftmargin=*]
\item If $\sigma_0,\sigma_1$ are from the query $\gsig(k_0,\eid,m_0)$ and $\gsig(k_1,\eid,m_1)$ for some $k_0,k_1\in \hu$ and $\gsk[k_0] = (x_{k_0},y_{k_0},A_{k_0} = (g_1 h^{-y_{k_0}})^{\frac{1}{\gamma+x_{k_0}}})$ and $\gsk[{k_1}] = (x_{k_1},y_{k_1},A_{k_1} = (g_1 h^{-y_{k_1}})^{\frac{1}{\gamma+x_{k_1}}})$. $\B$ computes $A_b = B_b D_b^{-\xi}$ for $b=0,1$. Since $\judge(\gpk,i_b,\upk[i_b],\sigma_b,\pi_b) \\= 1$, we know that $\pair(A_b, w g_2^{x_b'}) = \pair(g_1 h^{-y_{i_b}}, g_2)$ which means $A_b = (g_1 h^{-y_{i_b}})^{\frac{1}{\gamma + x_b'}}$. Because $\sigma_b$ is produced by $\gsig$ query, we have $A_b = (g_1 h^{-y_{i_b}})^{\frac{1}{\gamma + x_b'}} = A_{k_b} = (g_1 h^{-y_{k_b}})^{\frac{1}{\gamma + x_{k_b}}}$ for $b=0,1$. Replacing $g_1, h$ with $g^a, g^{az}$ correspondingly in the equation, we can obtain
$g^{\frac{1-z y_{i_b}}{\gamma+x_b'}} = g^{\frac{1-z y_{k_b}}{\gamma+x_{k_b}}}$ and thus ${\frac{1-z y_{i_b}}{\gamma+x_b'}} = {\frac{1-z y_{k_b}}{\gamma+x_{k_b}}}$. Then we have $x_{k_b}-x'_b = z[y_{i_b}(\gamma+x_{k_b})-y_{k_b}(\gamma+x'_b)]$.  Suppose $x'_b = x_{k_b}$ for $b=0,1$, from $A_b = A_{k_b}$, we can easily see  $y_{i_b} = y_{k_b}$. If $i_0 = i_1$, then $y_{i_0} = y_{i_1}$ and thus $T_0 = \hs(\eid)^{y_{k_0}} = T_1 = \hs(\eid)^{y_{k_1}}$. This means the $\link(\eid,m_0,\sigma_0,m_1,\sigma_1) = 1$. If $i_0 \neq i_1$, then $y_{i_0} \neq y_{i_1}$ and $\link(\eid,m_0,\sigma_0,m_1,\sigma_1) = 0$. This contradicts the hypothesis that $\A$ breaks the e-linkability. Therefore we can deduce that $x'_{b}\neq x_{k_b}$ for some $b\in\setd{0,1}$, then we find a solution $z = \frac{x_{k_b}-x'_b}{y_{i_b}(\gamma+x_{k_b})-y_{k_b}(\gamma+x'_b)}$ for the DL instance.
\item If $\sigma_0$ is from the query $\gsig(k_0,\eid,m_0)$ for some $k_0\in \hu$ and $\gsk[k_0] = (x_{k_0},y_{k_0},A_{k_0} = (g_1 h^{-y_{k_0}})^{\frac{1}{\gamma+x_{k_0}}})$, while $\sigma_1$ is not obtained from the $\gsig$ query. Similarly as above analysis, we have $A_0 = (g_1 h^{-y_{i_0}})^{\frac{1}{\gamma + x_0'}} = (g_1 h^{-y_{k_0}})^{\frac{1}{\gamma + x_{k_0}}}$. $\B$ then rewinds $\A$ on $\sigma_1$ to extract $A_1 = (g_1 h^{-\Delta y})^{\frac{1}{\gamma+\Delta x}}$. Because the $\judge$ is successful, then similarly we have $A_1 = (g_1 h^{-y_{i_1}})^{\frac{1}{\gamma+x_1'}}$ for some $x_1'$. Similarly as above, suppose $x_0'=x_{k_i}$ and $\Delta x = x_1'$, then we can deduce that $y_{i_0} = y_{k_0}$ and $\Delta y = y_{i_1}$. This means $\link$ outputs 1 when $i_0 = i_1$ and 0 when $i_0\neq i_1$. However, this contradicts to the hypothesis that $\A$ breaks the link-unforgeability. Therefore  $x_0'\neq x_{k_i}$ or $\Delta x \neq x_1'$. When one of the two inequations holds, we can use the similar method described above to obtain a solution $z$ for the DL instance.
\item The other cases are similar and are omitted here.
%\item The case when $\sigma_1$ is from $\gsig$ queries and $\sigma_0$ is not, and the case when $\sigma_0,\sigma_1$ are both not from $\gsig$ queries are similar as above and the details are omitted here.
\end{itemize}

\end{proof}

\end{document}